\documentclass{aa}  

\usepackage{graphicx}
\usepackage{txfonts}
\begin{document}

   \title{The Gaia-ESO Survey: Chemical tagging in the thin disk\thanks{Based on observations collected with the FLAMES instrument at
VLT/UT2 telescope (Paranal Observatory, ESO, Chile), for the Gaia-
ESO Large Public Spectroscopic Survey (188.B-3002, 193.B-0936, 197.B-1074).}}

   \subtitle{Open clusters blindly recovered in the elemental abundance space}

\titlerunning{Chemical tagging} 
\authorrunning{L. Spina et al.}

   \author{L. Spina\inst{\ref{oapd}}, L. Magrini\inst{\ref{oaa}}, G. G. Sacco\inst{\ref{oaa}}, G. Casali\inst{\ref{difa}, \ref{oas}}, A. Vallenari\inst{\ref{oapd}}, G. Tautvai{\v s}ien{\.e}\inst{\ref{vil}}, F. Jim\'{e}nez-Esteban\inst{\ref{CSIC}}, G. Gilmore\inst{\ref{cam}}, S. Randich\inst{\ref{oaa}}, S. Feltzing\inst{\ref{lund}}, R.~D. Jeffries\inst{\ref{keele}}, T. Bensby\inst{\ref{lund}}, A. Bragaglia\inst{\ref{oas}}, R. Smiljanic\inst{\ref{war}}, G. Carraro\inst{\ref{unipd}}, L. Morbidelli\inst{\ref{oaa}}, S. Zaggia\inst{\ref{oapd}}}

   \institute{INAF - Padova Observatory, Vicolo dell'Osservatorio 5, 35122 Padova, Italy\label{oapd}\\
              \email{spina.astro@gmail.com}
              \and
              INAF - Osservatorio Astrofisico di Arcetri, Largo E. Fermi 5, 50125, Firenze, Italy \label{oaa} 
              \and
              Dipartimento di Fisica e Astronomia, Universit\`a degli Studi di Bologna, Via Gobetti 93/2, I-40129 Bologna, Italy\label{difa}
              \and
              INAF–Osservatorio di Astrofisica e Scienza dello Spazio, Via P. Gobetti 93/3, 40129 Bologna, Italy \label{oas}
              \and
              Institute of Theoretical Physics and Astronomy, Vilnius University, Sauletekio av. 3, 10257, Vilnius, Lithuania \label{vil}
              \and
              Departamento de Astrof\'{i}sica, Centro de Astrobiolog\'{i}a (CSIC-INTA), E-28692 Villanueva de la Ca$\tilde{\rm n}$ada, Madrid, Spain \label{CSIC}
              \and
              Institute of Astronomy, University of Cambridge, Madingley Road, Cambridge CB3 0HA, United Kingdom \label{cam}
              \and
              Lund Observatory, Department of Astronomy and Theoretical Physics, Box 43, SE-221 00 Lund, Sweden\label{lund}
              \and
              Astrophysics Group, Keele University, Keele, Staffordshire ST5 5BG, United Kingdom\label{keele}
              \and
              Nicolaus Copernicus Astronomical Center, Polish Academy of Sciences, ul. Bartycka 18, 00-716, Warsaw, Poland\label{war}
              \and
              Dipartimento di Fisica e Astronomia, Universit\`a di Padova, Vicolo dell'Osservatorio 3, 35122 Padova, Italy\label{unipd}
             }

   \date{Received February 11, 2022; accepted March 28, 2022}


  \abstract
 {The chemical makeup of a star provides the fossil information of the environment where it formed. Under this premise, it should be possible to use chemical abundances to tag stars that formed within the same stellar association. This idea - known as \textit{chemical tagging} - has not produced the expected results, especially within the thin disk where open stellar clusters have chemical patterns that are difficult to disentangle.}
 {The ultimate goal of this study is to probe the feasibility of \textit{chemical tagging} within the thin disk population using high-quality data from a controlled sample of stars. We also aim at improving the existing techniques of \textit{chemical tagging} and giving some kind of guidance on different strategies of clustering analysis in the elemental abundance space.}
   {Here we develop the first blind search of open clusters' members through clustering analysis in the elemental abundance space using the OPTICS algorithm applied to data from the Gaia-ESO survey. First, we evaluate different strategies of analysis (e.g., choice of the algorithm, data preprocessing techniques, metric, space of data clustering), determining which ones are more performing. Second, we apply these methods to a data set including both field stars and open clusters attempting a blind recover of as many open clusters as possible.}
   {We show how specific strategies of data analysis can improve the final results. Specifically, we demonstrate that open clusters can be more efficaciously recovered with the Manhattan metric and on a space whose dimensions are carefully selected. Using these (and other) prescriptions we are able to recover open clusters hidden in our data set and find new members of these stellar associations (i.e., escapers, binaries).}
   {Our results indicate that there are chances of recovering open clusters' members via clustering analysis in the elemental abundance space, albeit in a data set that has a very high fraction of cluster members compared to an average field star sample. Presumably, the performances of chemical tagging will further increase with higher quality data and more sophisticated clustering algorithms, which will likely became available in the near future. }
   {}


   \maketitle
%

\section{Introduction}
The idea of chemical tagging originated about 20 years ago from the work of \citet{freeman02}, in which it was suggested that destroyed clusters could be found by means of their unique chemical signatures. Thus, the global star formation history of the Galactic disc could be reconstructed, also detecting possible debris of satellite systems.
The first principle of the theory is based on the observational evidence of the chemical homogeneity of open clusters and moving groups \citep{desilva06,  sestito07, bovy16, Poovelil20}. 
The second axiom is about the uniqueness of the chemical pattern of each cluster, and therefore the possibility, using a sufficient number of elements, to find stars born in the same cluster, although it no longer exists \citep{Bland-Hawthorn10}. Thus, the feasibility of chemical tagging relies on the presence and magnitude of cluster-to-cluster differences \citep[inter-cluster, see][]{Reddy20} and on the intra-cluster homogeneity. 
The chemical tagging based only on the use of chemical information is often called {\em strong} chemical tagging, and aims at recovering each unique site of star formation. 
A {\em weak} version of the chemical tagging aims at recovering broader stellar populations, like the thin and the thick discs, the halo, and their substructures \citep{Martell10, Hawkins15, Wojno16, Schiavon17, Recio-Blanco17}.

Progress in recent years has shown that reality is more complex than anticipated. 
On the one hand, the advent of large spectroscopic surveys, such as {\em Gaia}-ESO \citep{Gilmore12, Randich13}, APOGEE \citep{Majewski17} and GALAH \citep{DeSilva15}, which have devoted significant fractions of their time to observe star clusters, has shown that there can be measurable differences in the abundances of members of clusters in different phases. Those differences can be due to physical reasons, as, e.g., diffusion \citep[see, for instance][]{BertelliMotta18, Souto18, Liu19}, internal dredge-up \citep[e.g][]{Charbonnel95, Charbonnel07, Lagarde12}, planet engulfment events \citep[e.g][]{Tucci19,Spina15,Spina18,Spina21}, or due to analysis techniques \citep[see, e.g.][]{Casali20b}. 

On the other hand, the question of the uniqueness of the chemical pattern of each cluster is still a matter of debate \citep[see, e.g.][]{BlancoCuaresma15, PriceJones18}. The most plausible way to test it is definitely by trying to re-find members of known clusters, mixed between themselves and with stars belonging to the field.
In addition, the uniqueness is closely linked to the elements chosen to characterise the chemical pattern. Indeed, not all elements work equally well: in particular those produced at short scale times by massive stars, such as the $\alpha$ elements, and that have enriched the disk mostly in remote times, vary in similar way in all open clusters, making them of little use as diversity tracers \citep[see. e.g.][]{Ness21}.
On the other side, elements released only recently in the interstellar medium (ISM) by low-mass stars and with a strong metallicity-dependence of their yields \citep[see, e.g.][]{Casali20,Magrini21}, such as slow neutron capture elements, contribute differently to the chemical composition of the clusters of various ages and located in diverse parts of the disc, and might be better tracers. 

There have been several attempts in recent years to recover known stellar associations through clustering analysis in the elemental abundance space or to find the birth sites of stars born in disrupted open clusters. These works have shown disadvantages and advantages of the various types of approach \citep[see, e.g.][]{Mitschang14, Ting15, Ness18}. 
However, some works showed encouraging results. \citet{Hogg16} were able to identify (by the k-means algorithm) known globular clusters (such as M13 and M5), stellar streams, and other undetermined overdensities of stars in the 15-dimensional chemical-abundance space delivered by The Cannon \citep{Ness16}. Globular clusters were also recovered by \citet{chen18} with the Density-Based Spatial Clustering of Applications with Noise (DBSCAN; \citealt{Ester96}) algorithm. However, while \citet{Hogg16} only used the chemical abundance information, \citet{chen18} carried out the clustering analysis in a chemical-velocity space. Furthermore, both these works are not ``blind'', meaning that known members of stellar clusters were used to tune the parameters of the clustering algorithms. Finally, \citet{Hogg16} highlighted that it is easier for clustering analysis to find structures at low metallicity (e.g., in the Galactic halo) than at solar metallicity (e.g., the Galactic disk). In fact, abundance-space features are more prominent at the edges of the distribution than in the center \citep[see. e.g.][]{Ting16}.

Nevertheless, strong chemical tagging has been tested in the Galactic disk as well. \citet{Blanco-Cuaresma18} performed a phylogenetic analysis to recover members of known star clusters, reconstructing most of the original open clusters using differential abundances with respect to M~67, and discussing the best set of elements to recover clusters. \citet{PriceJones19} identified groups of stars in a synthetic set of birth clusters using {\sc DBSCAN} and their percentage of clusters with more than 10 members recovered is 40\% using only abundances. In a following paper, \citet{PriceJones20} repeated the experiment on the APOGEE data set identifying in the abundance space 21 candidate stars clusters, but they were not able to recover any known open cluster. \citet{GarciaDias19} carried out a test on eight clustering algorithms and with four dimensionality reduction strategies applied on a data set composed by members of both open and globular clusters. This work shows how different data analysis strategies must be carefully evaluated before attempting chemical tagging. Finally, \citet{Casamiquela21}  used abundances of member stars in open clusters from high-resolution spectroscopy to recover them with {\sc HDBSCAN}: they could recover about 30\% of the analysed clusters in groups containing at least 40$\%$ of the actual cluster members. Notably, these attempts of chemical tagging in the thin disk are non-blind searches of stellar associations within a data set exclusively composed by known members of open clusters. The unique exception in the literature is the study from \citet{PriceJones20}, which used a sample of 182,538 stars belonging to all the Galactic components. Their inability of recovering any of the known open clusters present in their data set clearly demonstrates the difficulty of the task. This is especially true when noisy data significantly blur the clusters we wish to recover. In fact, \citet{PriceJones20} noted how real members of known open clusters have a larger dispersion in the abundance space than those of the 21 groups they identified.

\subsection{Outline of the paper}

Inspired by the study of \citet{PriceJones20}, in the present work we devise an experiment that simulates a realistic attempt of strong chemical tagging in the thin disk, attempting to improve the existing methodology and giving some kind of guidance. Our full data set is composed by both field stars and members of 40 open clusters. This data set - described in Section~\ref{Sec:dataset} - is smaller than the APOGEE sample used by \citet{PriceJones20}, but it constitutes a very controlled set of abundances derived through higher-resolution spectra.

In the first part of the paper (Section~\ref{Sec:Methods}), we use a subset of our data composed exclusively by members of 13 open clusters. The overarching aims of this Section are i) considering all the most common methods used so far to tackle the problem of chemical tagging, ii) identifying their pros and cons, ii) ultimately selecting the one we consider to be the most appropriate for our problem.  Firstly, we discuss the rationale behind the choice of the {\sc OPTICS} algorithm and behind important preprocessing techniques of our data set. Another important point we address is that of the choice of metric, which is often overlooked, assuming the Euclidean one \citep{Hogg16, Blanco-Cuaresma18, PriceJones20, Casamiquela21}, but which can give important improvements in the identification of stars belonging to common formation regions. The choice of the Manahattan metric, which ensures the best contrast in distance between the different points, is perfectly suited to highlight the contrast between clusters and field stars. Then, we focus on defining the abundance space, by tackling the issue of number of elements vs quality of measurement, and especially the choice of elements based on their uniqueness in terms of nucleosynthesis channels \citep[see, e.g.][]{adi15, Blanco-Cuaresma18}. Initial work on chemical tagging considered that between 10 and 15 elements were needed to guarantee uniqueness across populations \citep{Bland-Hawthorn04, desilva07}. We aim at demonstrating with our experiment that we can have some advantage in reducing the overall dimensionality to only the strongest variant dimensions. As already pointed out by \citet{Mitschang14}, one might think that increasing dimensionality would bring improvements. However, in practice, adding elements that are difficult to measure (with few or weak lines, uncertain atomic data, etc.) or elements which duplicate the information since they have a common origin, can only lead to increased uncertainties and scatter. 

In the second part of the paper (Section~\ref{Sec:chemical_tagging}) we apply the most performing strategies of data analysis to the full data set. First we identify the best set of algorithm's hyper-parameters that maximise the number of stellar associations that are recovered in the abundance space among the 13 open clusters analysed in Section~\ref{Sec:Methods}. The resulting model is then applied to the rest of the data set with the aim of recovering the other known open clusters and other eventual groups of stars of potential interest. With this approach, the algorithm is going to identify in the abundance space a number of clusters, some of which are real stellar associations, while others are just spurious overdensities. Thus, we further investigated only the groups with the highest density in the space of orbital actions. Interestingly, a few of these latter coincide with open clusters that are blindly recovered through our analysis. We also recover new stellar members of open clusters that were lost from the membership analysis. Finally, we identify groups of stars clustering both in the abundance space and in orbital actions, which should be further investigated with follow-up observations.

Finally, in Section~\ref{Sec:conclusions} we provide a summary of our main results and we discuss future perspectives of strong chemical tagging.



In order to avoid confusion with the terminology, in the following sections we will refer to ``open cluster'' or ``stellar association'' as an ensemble of stars sharing the same origin and formation site, regardless of whether they are still physically bound or dispersed. Instead, with ``cluster'' or ``group'' we will refer to a group of stars identified by the clustering algorithm in the elemental abundance space, regardless of whether they are all part of a real ``stellar association'' or not.

\section{Data set}
\label{Sec:dataset}
In this paper we make use of the sixth internal data release (iDr6) of the {\em Gaia}-ESO survey \citep{Gilmore12, Randich13}. More specifically we use the catalog of the targets observed in the high resolution mode with the spectrograph UVES (resolving power, R=47,000, and spectral range 480.0-680.0~nm).  
The catalog includes parameters and abundances for 6,917 stars. From this initial sample we only consider the stars with a radial velocity determination, and analysed by the working group (WG) in charge for the analysis of stars with spectral type F-G-K, WG~11 \citep[see][]{Smiljanic14}. The selected targets are identified with \texttt{REC\_WG} equal to ``WG11''. Furthermore, in the data set that we are going to analyse in this paper we only include evolved stars as their abundances are not affected by either atomic diffusion \citep{Souto18,BertelliMotta18} or by planet engulfment events \citep{Spina21}, therefore we select only the stars with surface gravity $\log~g\leq 3.5$ dex and effective temperature $4500< T_{\rm eff}\leq 5500$ K. Also, given that for our analysis we require a controlled sample of very high-quality abundance determinations, we only consider the stars with microturbulence $\xi<2.0$ km s$^{-1}$, with the number of nodes that provided estimations of iron abundances \texttt{N\_FEH} $>$ 1, and signal to noise ratio \texttt{SNR} $\geq$50 pixel$^{\rm -1}$. In fact, chemical abundances of these stars are more reliable than those derived for the rest of the sample. Since our study is focused on the thin disk population, we consider only the stars with a distance above the Galactic midplane |z|$<$1.5~kpc (derived  using {\em Gaia} {\sc edr3} distances). In addition to that, we only include stars having abundance determinations of all the following elements: Fe~I, Na~I, Mg~I, Al~I, Ca~I, Sc~II, Ti~I, V~I, Cr~I, Mn~I, Co~I, Ni~I, Cu~I, Zn~I, Ba~II, Y~II, Nd~II, and Eu~II. Finally, we discarded the stars  whose parameters had convergence problems in at least one of the analysis nodes. We also discarded the entries for which  one node provided results which differs from the all Nodes mean by more than 500 K in T$_{\rm eff}$, or 0.5 dex in log g, or 0.5 dex in [Fe/H]. The list of these low quality results is provided in an internal report of {\sc wg11}, and are flagged in the final table. Thus, our final sample is composed of 937 stars that fulfill all these criteria.

Our data set still contains a number of stars in globular clusters. Since our experiment is about chemical tagging in the thin disk, we want to remove such objects. Therefore, assuming the membership probabilities P$_{GCs}$ from \citet{Vasiliev21} we reject all the stars with P$_{GCs}>$0.9. This reduces the data set to 800 stars which are either in open clusters or in the field.

With this study we want to evaluate our capability of recovering stars in open clusters hidden in our data set. Therefore, before starting with the clustering analysis, it is of fundamental importance to establish the membership of stars to these associations. In a recent study, \citet{Jackson22} made use of the Gaia astrometric solutions and {\em Gaia}-ESO radial velocities to determine 3D membership probabilities P$_{3D}$ of several stars in our data set. When available, here we use these membership probabilities to infer whether or not each single star is member of an open cluster. When these probabilities are not available, which is the case for clusters with few observed stars, we use  membership probabilities computed as in \citet{Magrini21} that consider as members stars within 2-$\sigma$ from the average radial velocity, proper motion and parallax of the cluster. The first from {\em Gaia}-ESO iDR6 and the others from {\em Gaia} {\sc edr3}.   
Given these probabilities, we consider members of open clusters all the stars with P$\geq$0.9. From these cluster members we discard the stars in clusters younger than 500 Myr. This is necessary because chemical abundances in younger clusters suffer of low accuracy as their metal content appears significantly lower than what it actually is depending on the level of stellar chromospheric activity \citep{YanaGalarza19,Baratella20,Baratella21,Spina20,Zhang21}. Therefore, after this further selection we count 306 stars in open clusters. There are 40 open clusters in our data set. Among these, 13 include at least 10 stars: these cluster members will be used to choose the best strategies of data preprocessing and clustering analysis (see Section~\ref{Sec:Methods}) and to select the best set of hyperparameters for the algorithm (see Section~\ref{Sec:search_param}). Thus, hereafter we will refer to this sample that counts 190 stars as the \textit{optimization} sample. In addition to that, there are 23 open clusters with a number of members ranging between 2 and 9. These are the clusters that are hidden in our data set and that we want to ``recover'' through our analysis. Hereafter we will refer to this sample of 112 stars as the \textit{hidden clusters} sample. Finally, there are four open clusters with only one member. For practical reasons these four stars will be considered as field stars along with all the other stars without an estimation of membership probability to any cluster or with P$\leq$0.1. The resulting sample of \textit{field stars} counts 456 objects. All the stars with membership probability P$\in$(0.1,0.9) are excluded from our data set. The removal of stars with an indefinite membership is necessary to maximise the accuracy of the metrics used to evaluate our ability to recover open clusters through chemical tagging.

\section{Methods}
\label{Sec:Methods}
The search for clusters in a multidimensional space relies on some fundamental decisions that have to be taken. These decisions, which can significantly improve or deteriorate the efficiency of chemical tagging \citep[see][]{GarciaDias19}, are about the choice of the clustering algorithm, the criteria to set its hyperparameters, the techniques of data preprocessing, and the choice of the space of parameters that needs to be analysed. Unfortunately there is no a general method of data clustering nor a rule of thumb that ensure good results for a reasonably broad range of cases or situations. Instead, the best course of actions is strongly influenced by the data set one has at hand and by the type of cluster one aims to recover. For that reason it is necessary to think very carefully to each step of the analysis and evaluate whether or not it can boost the performances of the clustering algorithm.

In this Section we want to explore different approaches for clustering in the abundance space. Our aim is to identify the empirical and conceptual criteria that should drive our choices, and to outline - when possible - general strategies that can improve the efficiency of chemical tagging.

\subsection{The algorithm}
\label{Sec:algorithm}
\textit{Cluster} is a tricky concept in data science which varies depending on the type of problem that is tackled. For instance, a data set may be formed only by clusters or instead it may contain clusters plus scattered data. Clusters may have the same density profiles or be significantly different from each other. In some cases we may also be interested in detecting sub-clusters within the same agglomerate of data points. Given all this diversity, various techniques of cluster searches have been developed and several clustering algorithms are now available to us. Clusters found by one algorithm will definitely be different from clusters found by another algorithm, however there is no best algorithm for clustering analysis because their outcomes can be evaluated with very different criteria and under various points of view. The best way to get oriented in the vast panorama of machine learning algorithms is always understanding the type of problem that we need to solve and the data set that we have at hand. 

First, for practical reasons we restrict the choice of the algorithm to the several clustering techniques that are already implemented in  {\sc Python}. Second, our specific problem - chemical tagging - implies the search of clusters in a data set containing points that do not belong to any agglomerate. Thus, we must exclude the algorithms that are unable to process scattered or noisy instances, such as affinity propagation, agglomerative clustering, K-means, mini-batch K-means, spectral clustering, Gaussian mixing models, {\sc Ward}, and the balanced iterative reducing and clustering using hierarchies (BIRCH)\footnote{For more details on these different methods of clustering analysis, see \citealt{Rokach08}.}. Third, as we will demonstrate in Section~\ref{Sec:chemical_space}, open clusters can have different densities in the abundance space when they are seen by large spectroscopic surveys. Therefore, we must also exclude {\sc DBSCAN} as it assumes that all clusters have the same density. This leaves our choice to two algorithms: {\sc OPTICS} and {\sc HDBSCAN}.

The philosophy behind {\sc OPTICS} and {\sc HDBSCAN} is very similar. Both these algorithms work by detecting areas where points are concentrated more than other areas that are empty or sparse. These algorithms start the analysis with the first point of the data set and keep expanding to the neighbours points. While considering a specific point in a data set, the algorithm assigns to it a value called \textit{reachability distance}, which can be understood as a measure of the average density of the \textit{neighbourhood} around the point under consideration. The complete description of {\sc OPTICS} and {\sc HDBSCAN} and how they work is beyond the scope of this paper, but it can be found in the seminal papers by \citet{Ankerst99,Campello13}. What is important to highlight here is that both {\sc OPTICS} and {\sc HDBSCAN} do not produce a clustering of a data set explicitly, but instead they create an augmented ordering of the data points representing its density-based clustering structure. In practice, all the points of the data set are (linearly) ordered such that spatially closest points become neighbors in the new linear ordering. In addition, the \textit{reachability distance} assigned to each point gives a measure of the density of its \textit{neighbourhood}. Therefore, since a cluster can be defined as a region of high density separated by regions of low density, two points that are adjacent in the new linear ordering and that have very different \textit{reachability distances} can be seen as the ``beginning'' or the ``end'' of a cluster. That explains why the new indices of the reordered data set and the \textit{reachability distance} are the fundamental information used by both {\sc OPTICS} and {\sc HDBSCAN} to extract clusters and sub-clusters. Once this information is reached, the two algorithms differ in the technique used to identify these agglomerates.

For our analysis we decide to use the {\sc OPTICS} algorithm. This choice is driven by a few reasons. First, the {\sc OPTICS} code that is implemented in {\sc Python} as part of the \texttt{scikit-learn} package allows the user to visualize in a very simple and intuitive way the information that is at the basis of the cluster extraction. This information is fully represented in the {\em reachability plot} that shows the reachability distance of every point in the database - with the exception of the starting point for which the reachability distance is not defined - as a function of the new index assigned by the algorithm (see an example in Fig.~\ref{Fig:standardization}). The clusters are identified as the valleys or dips in the reachability plot, while the saddles are cluster boundaries or noise. Thus, the reachability plot represents a stand-alone tool to get key insights into the distribution of a data set, very useful to evaluate how different data preprocessing steps and different strategies for clustering analysis can improve (or not) the efficiency of the clustering algorithm. This is a very precious piece of information because, as we will see in the following sections, the many insights that we will learn from the reachability plots are directly applicable also to {\sc HDBSCAN} and other algorithms as well. Furthermore, the {\sc OPTICS} algorithm is simpler than HDBSCAN. As we will show in Section~\ref{Sec:evaluation_metrics}, {\sc OPTICS} mostly relies on one important hyperparameter that is unknown \textit{a priori}. Instead {\sc HDBSCAN} has more parameters that have to be tuned. Finally, {\sc HDBSCAN} has been already used in other studies \citep[e.g.][]{GarciaDias19,Casamiquela21}. Instead, there are no experiments of chemical tagging with {\sc OPTICS}. Thus, we want to take advantage of the present study as an opportunity to fill this gap.

\subsection{The hyperparameters}
\label{Sec:hyperparameters}

The OPTICS algorithm relies on the following hyperparameters:

\begin{itemize}
    \item \texttt{min\_samples}. This parameter is strictly related to the \textit{reachability distances}. In particular, it defines how large a \textit{neighbourhood} of a specific point is. A small \texttt{min\_samples} value will result in small \textit{neighbourhoods}, thus the resulting \textit{reachability distances} will be more sensitive to noise. Instead, a large  \texttt{min\_samples} value generates large \textit{neighbourhoods}, with a \textit{reachability distance} that is more stable against noise, but is also less sensitive to data sparsity. Therefore a ``good'' value of the \texttt{min\_samples} is the one that can capture the sparsity of our data set while filtering the noise out. This is the only hyperparameter that we do not set to a fixed value, as it is unknown \textit{a priori}. Every time that we use {\sc OPTICS}, we will search for the ``best''  \texttt{min\_samples} across the integer interval between 2 and 30. The ``best''  \texttt{min\_samples} value is determined accordingly to the evaluation metrics defined in Section~\ref{Sec:evaluation_metrics}.

    \item \texttt{max\_eps}. The maximum distance between two samples for one to be considered as in the neighborhood of the other. We assign $\infty$ to this parameter, so that {\sc OPTICS} can identify clusters across all scales.

    \item \texttt{metric}. Metric to use for distance computation. In our analysis we use the \textit{Manhattan} metric. The reason of this choice is explained in Section~\ref{Sec:metric}.
    
    \item \texttt{xi}. Determines the minimum steepness on the reachability plot that constitutes a cluster boundary. In other words, the \texttt{xi} parameter can be intuitively understood as a contrast parameter that establishes the increase in density expected for a cluster relatively to the noise. This parameter directly controls the number and the types of clusters we will obtain. A higher \texttt{xi} value can be used to find only the most significant clusters and it could be used to separate - for instance - the thin and thick disk populations in the chemical space. Instead, a lower \texttt{xi} value is better suited find less significant clusters, such as open clusters. Given that open clusters are the smallest building blocks of our Galactic disk and that they should be perfectly homogeneous except for the noise intrinsically present in the abundance determinations, we set \texttt{xi} to the smallest value that it can take. Namely, we empirically find that any \texttt{xi}$\leq$0.001 does not affect the results of the clustering analysis applied to our data set. Instead, any \texttt{xi}$>$0.001 can reduce the number of clusters found by the algorithm. Therefore we set \texttt{xi} = 0.001. 
    
    \item \texttt{min\_cluster\_size}. Describes the number of points required to form an {\sc OPTICS} cluster. Thus, the value of this parameter depends on the type of clusters the user is looking for. When evaluating the different strategies of clustering analysis, we will aim at extracting at least half of the members of open clusters composed by 10 members or more. Therefore, in this case we will set \texttt{min\_cluster\_size} = 5. Instead, in Section~\ref{Sec:chemical_tagging}, when we attempt the proper experiment of chemical tagging, the number of members of each stellar association is unknown, therefore we will set \texttt{min\_cluster\_size} = 2, which is the lowest value it can take.
    
\end{itemize}

\subsection{The evaluation metrics}
\label{Sec:evaluation_metrics}

The consequence of setting \texttt{max\_eps} = $\infty$ and \texttt{xi} = 0.001 is that {\sc OPTICS} finds a hierarchical structure of clusters encompassing from extremely large clusters that include almost all the points in our data set down to very small clusters composed only by few objects. The \texttt{min\_cluster\_size} = 5 filters out the clusters with less than 5 points. We also remove the clusters formed by $>$100 points. We define as N$_{g}$ the number of the remaining groups found by {\sc OPTICS}. Below we also define the evaluation metrics that we use to search the ``best'' \texttt{min\_samples} value and to assess different strategies of data analysis.

\begin{itemize}
         \item \texttt{Entropy}. Let $\{$idx$\}_i$ be the indices assigned by {\sc OPTICS} to the n$_{i}$ stellar members of the i$^{th}$ open cluster OC$_{i}$ (e.g., among the 13 open clusters of the \textit{optimization} sample). Let [1..n$_{i}$] be the integer interval between 1 and n$_{i}$. The entropy score of the i$^{th}$ open cluster is defined as 
         \begin{equation}
        E_i=\frac{\sigma(\{idx\}_i)-\sigma([1..n_{i}])}{\sigma(\{idx\}_i)+\sigma([1..n_i])},
        \end{equation}
        where $\sigma$ is the standard deviation. The entropy score ranges within [0,1]. A low E$_{i}$ value indicates that {\sc OPTICS} has effectively reorganized the data set in a way that members of OC$_{i}$ are neighbors in the reachability plot. Instead, a high E$_{i}$ value implies that OPTICS has not efficiently recognised the similarity of the members of OC$_{i}$. The median of the E$_{i}$ derived for all the open clusters defines the global entropy score E.
        
        \item \texttt{Homogeneity} and \texttt{completeness}. Given a group g found by {\sc OPTICS}, the homogeneity score is defined as
        \begin{equation}
        H_i=\frac{\rm number~of~OC_i~members~included~in~g}{\rm number~of~components~of~g}.
        \end{equation}
        Instead the completeness score is defined as 
         \begin{equation}
        C_i=\frac{\rm number~of~OC_i~members~included~in~g}{\rm number~of~members~of~OC_i}.
        \end{equation}
        These scores are derived only for the open clusters with H$_{i}$ and C$_{i}$ $\geq$ 0.5. These are the open clusters that - we say - have been recovered by {\sc OPTICS}. The median of these two scores of the open clusters for which they are defined respectively give the global homogeneity score H and the global completeness score C.

        \item \texttt{V measure}. It is the {\it harmonic} mean of two the desirable aspects of clustering: homogeneity and completeness. It is defined as it follows:
         \begin{equation}
        V_i=\frac{2 \times H_i \times C_i}{H_i + C_i}.
        \end{equation}
        The V measure is defined only for the open clusters with H$_{i}$ and C$_{i}$ $\geq$ 0.5. The median of V$_i$ gives the global V measure V.

        \item \texttt{Number of recovered OCs}. This is the number N$_{rec}$ of open clusters that have been recovered by {\sc OPTICS}. An open cluster OC$_{i}$ is "recovered" when both H$_{i}$ and C$_{i}$ are $\geq$ 0.5.
        
        \item \texttt{Clustering Efficiency}. This score is defined as the ratio between the number of clusters that have been recovered and the number of groups found by {\sc OPTICS}:

         \begin{equation}
        CE=\frac{\rm N_{rec}}{\rm N_{g}}.
        \end{equation}
\end{itemize}

Clustering algorithms are extremely sensitive to the choice of their hyperparameters. As we have discussed in Section~\ref{Sec:hyperparameters}, we fix them all with the only exception of \texttt{min\_samples}. This is a very delicate parameter that has a great impact on the final results. When \texttt{min\_samples} is too high, the algorithm does not detect any variation in density across the data set. As a consequence, the distribution of points in the reachability plot is flat, and the algorithm does not find any group. On the other hand, when \texttt{min\_samples} is too low,  the reachability plot would show a significant level of granularity. Therefore, the code would probably find the real clusters along with a myriad of other groups in the noise. Thus, tuning the \texttt{min\_samples} parameter to very low values would give the idea that the algorithm is able to recover a large fraction of clusters, however that may happen because the algorithms is generating several possible combinations of groups \citep[e.g.][]{Casamiquela21} and not because it is able to efficiently recognize patterns in the data set. It is not desirable a model that detects several clusters, many of which are noise and just a few are real. That model is unable to properly generalise when applied to a real data set. Therefore, in order to avoid overfitting, when evaluating the response of OPTICS to the variation of \texttt{min\_samples} we only consider the cases where CE is above a certain threshold. Among these cases alone we select the \texttt{min\_samples} value that maximises N$_{\rm rec}$. When the clustering analysis is applied to the $\textit{optimization}$ sample, we impose CE$>$0.5. In other words, we consider only the cases when more than half of the groups identified by {\sc OPTICS} correspond to real open clusters. Instead, when analysing the extended data set which also includes field stars and the \textit{hidden} clusters, it is reasonable to assume a lower threshold because the number of stellar associations that one can possibly recover is larger than the number open clusters that we use for the optimization (for more details see Section~\ref{Sec:chemical_tagging}).

\subsection{Reproducibility}
\label{Sec:index_ordering}

The distance- and density-based algorithms - such as k-means, {\sc DBSCAN}, {\sc HDBSCAN} and {\sc OPTICS} - are not entirely deterministic. The sequence of points in the linearly re-ordered data set, thus also the stars belonging to each group, slightly depend on how the data set is initially ordered. Therefore, to obtain results that are fully reproducible and objective it is strictly necessary to define a criterion that unequivocally pre-orders the data set before it is fed to the clustering algorithm. All that also implies that there must be a way to pre-order the data set that can make the algorithm more efficient than just analysing a random sequence of instances.

\begin{figure}
\centering
  \includegraphics[width=0.45\textwidth]{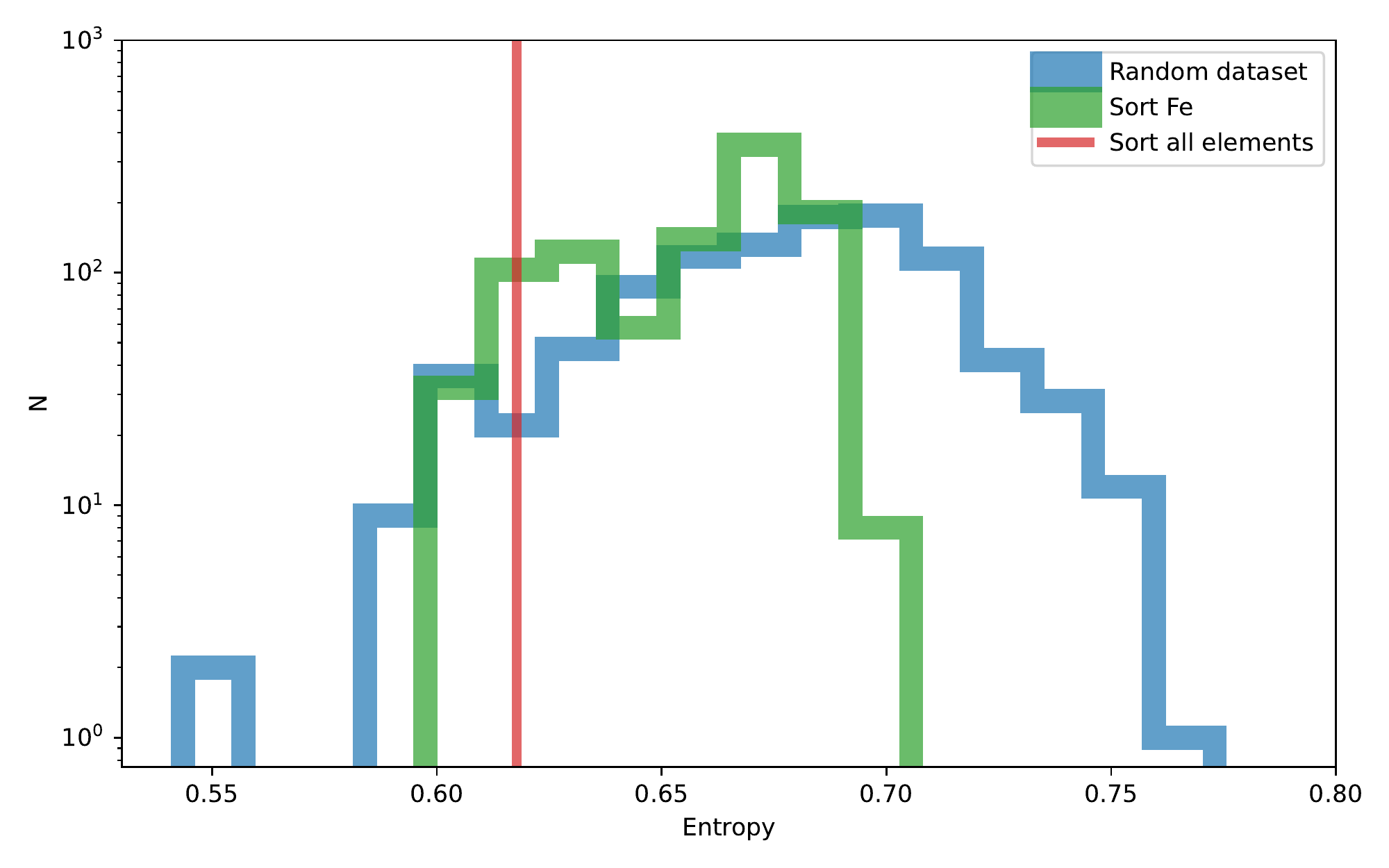}
  \caption{The entropy values resulting from the 1000 trials of clustering analysis carried out by {\sc OPTICS} on a data set that is randomly ordered (blue histogram), and on a data set that was sorted as a function of [Fe/H] (green histogram). Finally the red vertical line shows the entropy that is obtained when the data set is sorted as a function of the [X/H] abundances, starting with [Fe/H] followed by all the other elements.
  }
  \label{Fig:indexing}
\end{figure}

In Fig.~\ref{Fig:indexing} we show the distribution of the entropy coefficients obtained by repeating 1000 times the clustering analysis on a randomly ordered data set (in blue), and on a data set sorted as a function of [Fe/H] (in green). At each iteration, the clustering analysis is carried out following the recipe detailed in Section~\ref{Sec:evaluation_metrics} and using all the 19 elemental abundances as the input features.
When the data set is randomly ordered the resulting entropy spans a large range of values between 0.55 and 0.80. However, when we sort input data set as a function of Fe abundance, the entropy distribution is mostly restricted between 0.6 and 0.7. Finally, when we sort the data set as a function of [Fe/H] and of all the other abundances [X/H], we always obtain the same entropy of 0.62 (see red line). This last entropy is one of the lowest value that we have obtained across all the previous iterations. Therefore, pre-ordering the data set as a function of [X/H], starting with [Fe/H] and  followed by all the other elements, is a good strategy to improve the performances of the algorithm. In light of these results, we will apply this data preprocessing step in the following analysis.

\subsection{Standardization}
\label{Sec:standardization}

Standardization is a scaling technique in which a variable's distribution is transformed in order to have its mean equal to zero and a unit standard deviation. Data standardization makes data dimensionless. By standardizing one attempts to give all variables an equal weight, in the hope of achieving objectivity. In fact, if one feature has a range of values much larger than the others, clustering would be completely dominated by that single feature. Therefore, standardizing data is recommended because otherwise the range of values in each feature will act as a weight when determining how to cluster data, which is undesired in most cases. However, in practice standardization is not always strictly necessary.

\begin{figure*}
\centering
  \includegraphics[width=0.95\textwidth]{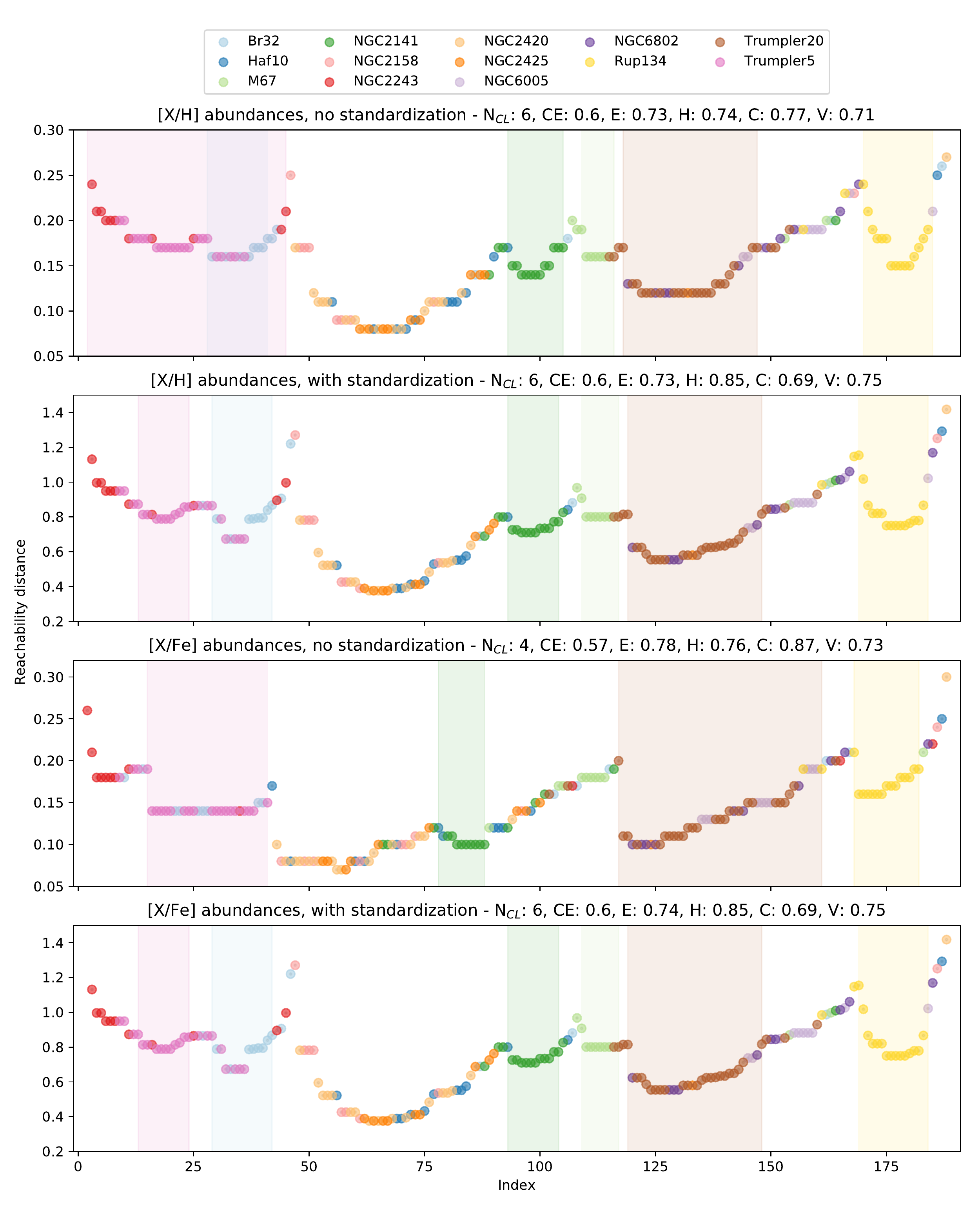}
  \caption{The panels show the reachability plots resulting from clustering with different strategies of standardization. The points are color coded depending on the cluster they are member of. Vertical colored strips indicate the groups associated to the recovered open clusters. The relevant coefficients and scores are listed on top of each panel. \textbf{Top panel.} Reachability plot obtained from clustering analysis in the [Fe/H]-[Si/H]-[Ni/H]-[Co/H]-[Ba/H] space without standardization. \textbf{Second panel.} Reachability plot obtained from clustering analysis in the [Fe/H]-[Si/H]-[Ni/H]-[Co/H]-[Ba/H] space with standardization. \textbf{Third panel.} Reachability plot obtained from clustering analysis in the [Fe/H]-[Si/Fe]-[Ni/Fe]-[Co/Fe]-[Ba/Fe] space without standardization. \textbf{Bottom panel.} Reachability plot obtained from clustering analysis in the [Fe/H]-[Si/Fe]-[Ni/Fe]-[Co/Fe]-[Ba/Fe] space with standardization. 
  }
  \label{Fig:standardization}
\end{figure*}

In Fig~\ref{Fig:standardization} we compare the effects of standardization in the analysis of our data set. Points are color coded depending on the open cluster they are member of. In the first panel, we show the reachability plot derived by {\sc OPTICS} which has searched for clusters in the abundance space [Fe/H]-[Si/H]-[Ni/H]-[Co/H]-[Ba/H]. No standardization is applied to the data set before the analysis. Instead, the second panel shows the reachability plot obtained through the same analysis, but on the standardised data set. On top of each panel we list the metrics and coefficients, while the colored bands highlight the clusters that have been recovered by {\sc OPTICS}. 

Solely based on the number of recovered open clusters and the other metrics and coefficients, standardization does not produce a significant improvement in the performances of the clustering algorithm. This is because all the features in the abundance space that we have used span similar ranges of values and have similar distributions. Therefore the {\sc OPTICS} will give them similar weights even without standardization. However, from a more careful analysis of the reachability plots it is possible to notice that standardization allows us to detect smaller density variations within each cluster. This is particularly evident for Trumpler~20, whose members mostly share the same reachability distance when no standardization is applied. Instead, when data are standardized the densest region of the cluster, corresponding to the points with a lower reachability distance, is more easily identifiable.

In order to make the benefits of standardization appear more evident we use the following features: [Fe/H]-[Si/Fe]-[Ni/Fe]-[Co/Fe]-[Ba/Fe]. In this case [Fe/H] spans a much larger range of values than all the other features. In fact, when we apply {\sc OPTICS} without any standardization, we obtain much worse results (see the third panel): the algorithms is able to recover only four open clusters instead of the six of the previous analysis. In this case standardization can significantly improve the efficiency of the algorithms (see the fourth panel).

In conclusion, although there are cases when standardization would not produce significant improvements in terms of the metrics that we are using to evaluate our results, this preprocessing technique will always positively contribute in finding overdensities in the abundance space. Therefore, it should always be applied even when features have very similar distributions. These conclusions apply for {\sc OPTICS}, and many other clustering algorithms (e.g., K-means, {\sc DBSCAN}, {\sc HDBSCAN}, etc...). 

Hereafter, we always standardize the data set before the clustering analysis.

\subsection{The metric}
\label{Sec:metric}

The metric is another critical ingredient in clustering analysis. The metric determines how distances between points are estimated. Therefore the contrast in distance between points in a dataset heavily depends on which metric is chosen. All previous experiments on chemical tagging through cluster analysis in the abundance space have used the Euclidean metric \citep[][]{Hogg16,Blanco-Cuaresma18,PriceJones20,GarciaDias19,Casamiquela21}, which is certainly the most familiar one and for that reason it could appear as the most natural choice. However, there are cases where the Manhattan metric may be preferable to Euclidean distance. In fact, the Euclidean metric is much more sensitive against to noise than the Manhattan metric. That is readily evident when considering how the two distances between points \textbf{p} and \textbf{q} is calculated. The Euclidean distance is defined as

\begin{equation}
d(\bf{p},\bf{q})_{\rm E}= \sqrt{\sum_i (p_i-q_i)^2},
\end{equation}

while the Manhattan distance is 

\begin{equation}
d(\bf{p},\bf{q})_{\rm M}= \sum_i |p_i-q_i|.
\end{equation}

Because of the square exponent, the Euclidean metric gives more emphasis to the outliers regardless of whether they are just noisy data or not. Instead, the Manhattan is more stable against noise. 

Certainly there are applications where a high sensitivity to noise is desirable, for instance when the features  are measured with a precision that is much higher than the typical separation of clusters. Unfortunately, this is not our case. Chemical abundances from spectroscopic surveys are obtained through automatic pipelines built to analyse, in a reasonable amount of time, thousands of spectra with very different signal-to-noise ratios and from stars having a broad range of atmospheric parameters, rotational velocities, and compositions. Thus, Galactic surveys typically prioritise large statistics over precision and accuracy. Therefore, given the unavoidable presence of noise, the Euclidean metric blurs clusters in the abundance space more than the Manhattan metric. On top of that, when we are working in a space that has more than a few dimensions - which is exactly our case - the Manhattan metric is the one that ensures the best contrast in distance between the different points \citep{Aggarwal01}, hence also the best density contrast between clusters and empty regions.

\begin{figure*}
\centering
  \includegraphics[width=0.95\textwidth]{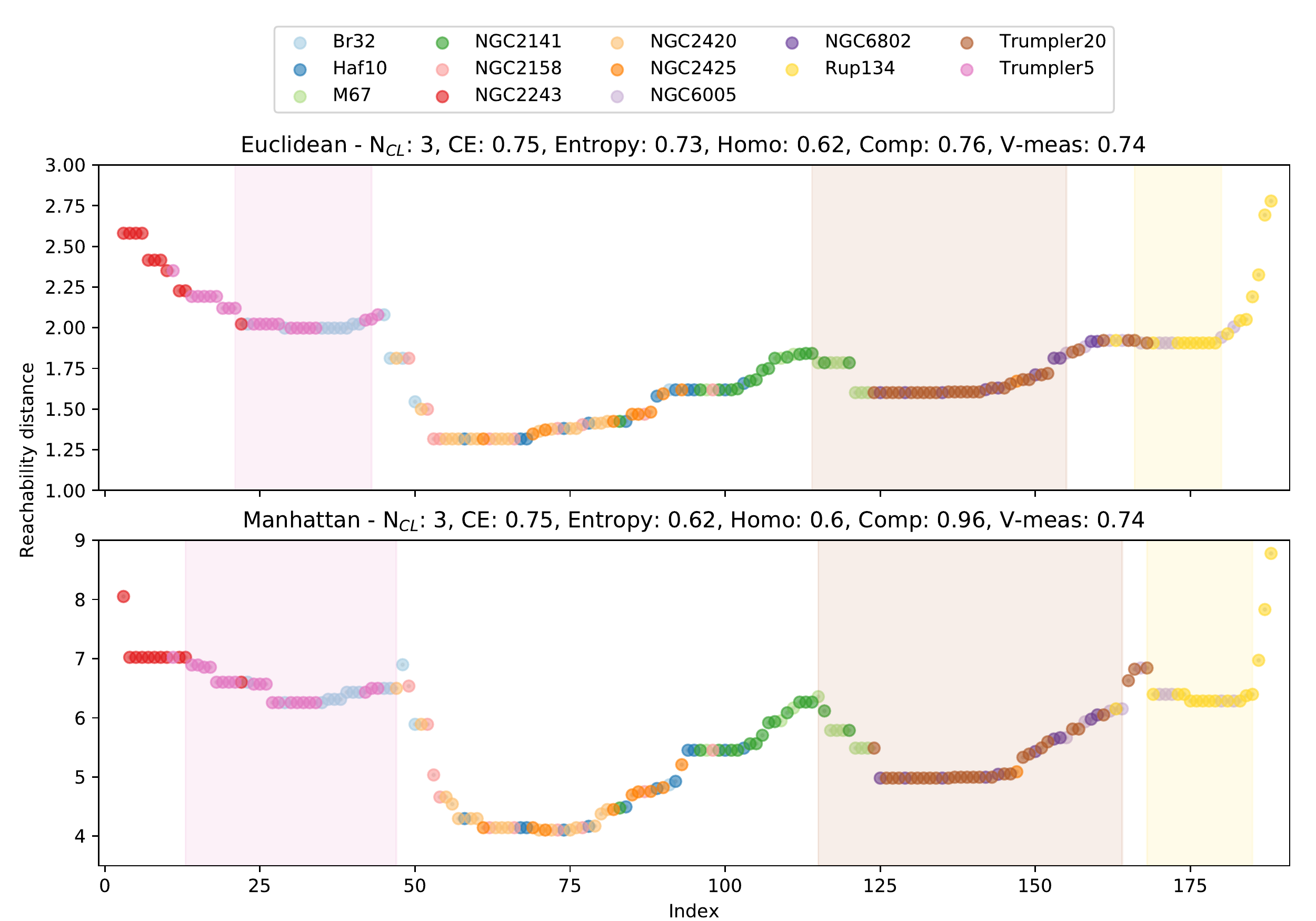}
  \caption{The panels show the reachability plots resulting from clustering with different metrics. The points are color coded depending on the cluster they are member of. The relevant coefficients and scores are listed on top of each panel. \textbf{Top panel.} Reachability plot obtained with the Euclidean metric. \textbf{Bottom panel.} Reachability plot obtained with the Manhattan metric.
  }
  \label{Fig:metric}
\end{figure*}

In Fig.~\ref{Fig:metric}, we compare the results obtained though the two different metrics. The abundance space considered here is defined by the [X/H] abundances of all the 19 elements (Fe, Na, Mg, Al, Si, Ca, Sc, Ti, V, Cr, Mn, Co, Ni, Cu, Zn, Ba, Y, Nd, and Eu). We notice some key differences between the outcomes of the two analysis. First, the Manhattan distance gives a smaller entropy score E, meaning that with this metric the data set is more efficiently ordered. This is expected as Manhattan is less affected by noisy data than the Euclidean metric. Second, the reachability distance in the Manhattan metric spans a wider range of values than in the Euclidean metric. As a result of this higher contrast in the distances between points, dips and valleys in the reachability plot obtained with the Manhattan metric will be deeper than those derived with the Euclidean distance. When dips and valleys are deeper, it is easier to identify clusters in the data set.

In conclusion, the Manhattan metric is a better option than the Euclidean for our specific application. This conclusion holds even for other distance- or density-based clustering algorithms, such as K-means, {\sc DBSCAN, HDBSCAN}, etc... Therefore, in the rest of this paper we will always use Manhattan distance.

\subsection{The elemental abundance space}
\label{Sec:chemical_space}

The definition of the elemental abundance space is one of the most important decisions that has to be taken before applying the clustering analysis. The option that seems the most obvious and natural is to use all the elemental abundances that one has in hand to define the space where to look for clusters. That strategy is the one typically used in the past \citep{Hogg16,chen18,PriceJones20,Casamiquela21}.

\begin{figure*}
\centering
  \includegraphics[width=0.95\textwidth]{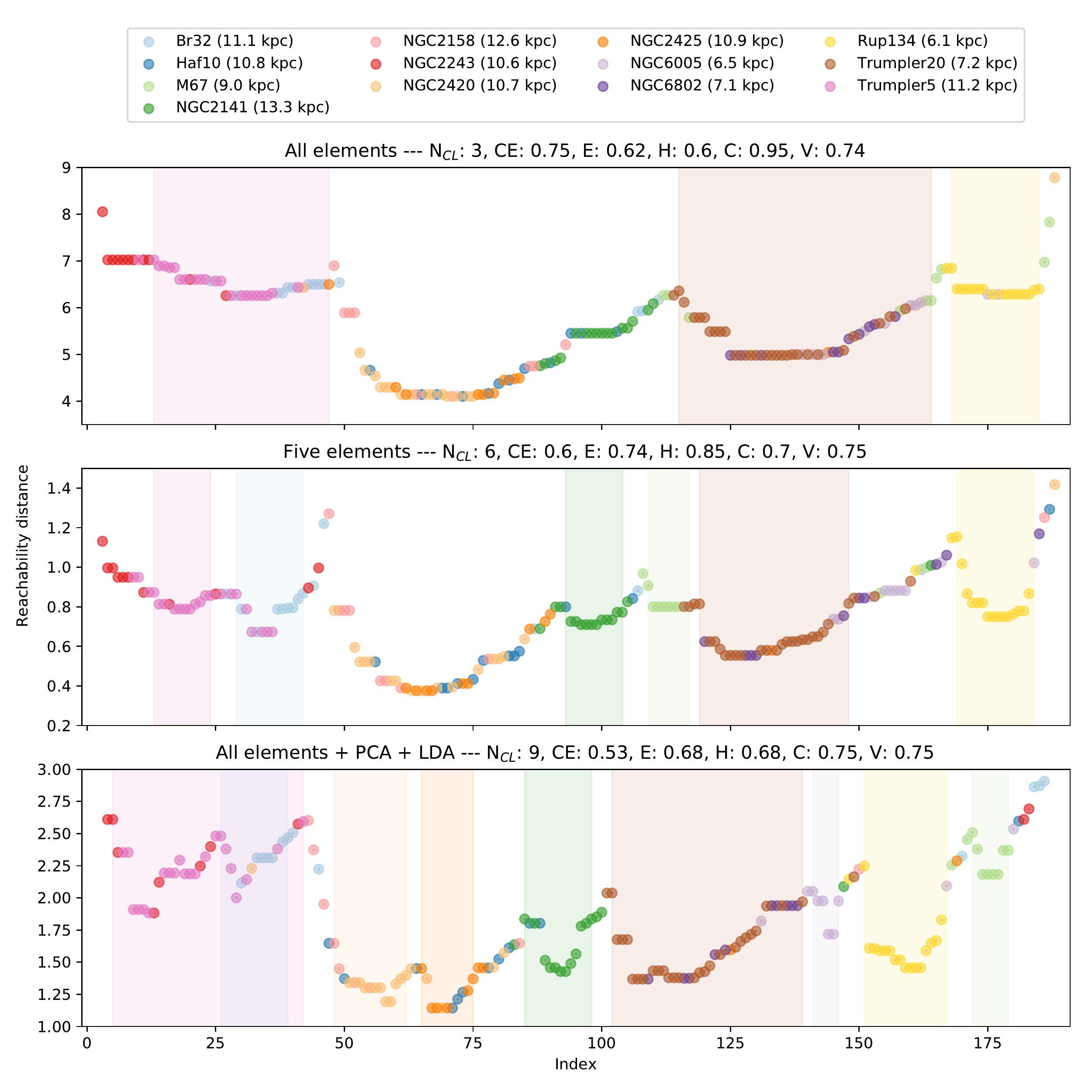}
  \caption{The panels show the reachability plots resulting from clustering on different abundance spaces. The points are color coded depending on the cluster they are member of. The relevant coefficients and scores are listed on top of each panel. \textbf{Top panel}. Reachability plot obtained from the 19-dimensional space defined by the chemical abundances of Fe, Na, Mg, Al, Si, Ca, Sc, Ti, V, Cr, Mn, Co, Ni, Cu, Zns, Ba, Y, Nd, and Eu. \textbf{Mid panel}. Reachability plot obtained from the 5-dimensional space defined by the chemical abundances of Fe, Si, Ni, Co, and Ba. \textbf{Bottom panel}. Reachability plot obtained from the 5-dimensional space found by PCA and LDA.
  }
  \label{Fig:elements}
\end{figure*}

The first panel in Fig.\ref{Fig:elements} shows the results obtained with all the 19 elemental abundances that have been measured in our sample. As we notice, the results are not entirely satisfactory as only three open clusters out of 13 have been recovered. Instead, when we decrease the dimensionality of the abundance space to the abundances [X/H] of only five elements (i.e., Fe, Si, Ni, Co, and Ba), the algorithm is able to recover up to six open clusters (see second panel). This significant difference in the performance is due to some common difficulties arising when attempting clustering analysis in a high-dimensional space. These issues are  known as ``the curse of dimensionality'' \citep{Bellman57}. 

First, when we increase the number of dimensions, our data becomes more sparse: every new dimension increases the volume of the abundance space, giving our data a higher differentiation chance. As a result, it is also more difficult to find groups of stars sharing a common chemistry. 

Second, as we increase the number of dimensions, the average distance between two points of our data set increases as well. Therefore, the relative distances between points get blurred. These two effects are visible by comparing the first two panels in Fig.~\ref{Fig:elements}. For instance, while the distribution of members of the cluster Ruprecht~134 is flat in the first panel, it clearly shows a dip in the second panel. Meaning that in a lower-dimensional space the algorithm is more efficient in recognizing overdensities.

Third, as we increase the number of dimensions, we likely introduce features that do not add much value to our model. When this is the case, the model will learn from noisy or irrelevant features. This can lead to a reduction in performance of the clustering algorithm, especially when abundances are derived with a limited precision such as in the case of those produced by large spectroscopic surveys.

On top of the ``the curse of dimensionality'', there are two other aspects that they should always be considered, although they do not affect algorithms' performance. Namely, working in a higher-dimensional space can significantly increase the time of the analysis and also lead to more complex models that are harder to interpret.

In conclusion, the abundance space defined by all the 19 elements contains a redundant number of dimensions. The simplest way to define a space with a smaller number of dimensions is to identify which elements that are the most meaningful among those that we have detected. To do so, we calculate for each element the Calinski-Harabasz score \citep{Calinski1974}, which is defined as the ratio of the sum of between-cluster dispersion and of within-cluster dispersion for all clusters. These scores are calculated on all stars belonging to the \textit{optimization} sample. The values are listed in the second column of Table~\ref{Table:Calinski}. The higher is the Calinski-Harabasz score and the more the [X/H] abundance can differentiate between members of different clusters. Note that this score is also influenced by the typical uncertainty of abundance determinations for each element. In fact, under the assumption that stellar association are chemically homogeneous, a small within-cluster dispersion implies a high precision in the abundance determination. In other words, a chemical element with very uncertain abundance determinations is expected to have a low Calinski-Harabasz even though it is highly distinctive of different stellar associations. Thus, the elemental abundances with the highest Calinski-Harabasz score are expected to be the most distinctive of stellar associations and the most precisely determined.

However, considering the elements with the highest Calinski-Harabasz score is not enough. In fact, two elements with an equally high score might have been produced from the same nucleosynthetic channel (e.g., see Mg and Si), therefore the inclusion of both of them would not add much information to the system \citep{Ness21}. Since each element has its own story to tell, in addition of using the Calinski-Harabasz score, we also want elements mostly produced though different nucleosynthetic processes \citep{Ting21}. The main nucleosynthetic origin of each element is inferred from \citet{Kobayashi20} and they are listed in the third column of Table~\ref{Table:Calinski}: Core Collapse supernovae (CC), Type Ia supernovae (Type Ia), HyperNovae (HNe), and Asymptotic Giant Branch stars (AGB). Given the CH score and the nucleosynthetic channel, we can select the five most meaningful elements among the 19 available: Si, Fe, Co, Ni, and Ba. By reducing the dimensions of the abundance space, from 19 to five, the performances of the {\sc OPTICS} algorithm significantly increase, as it is evident from Fig.~\ref{Fig:elements}-mid panel. Note that these five elements are not necessarily the best ones to recover the largest number of clusters from our data set, however, they certainly produce better results than just simply taking all the elements. Although our choice of the most meaningful elements is driven by some objective criteria, one could also adopt a more heuristic approach increasing or decreasing the number of features, testing different combinations of elements and empirically verifying how the result change. What we want to highlight here is just that, for our specific problem, we observe a significant improvement in performance when we select fewer more meaningful features.

\begin{table}
\caption{Criteria used to select the most meaningful elements}              
\label{Table:Calinski}      
\centering      
\begin{tabular}{ccc}      
\hline 
Element & CH score & Nucleosynthetic \\    
  &   & origin \\    
\hline

Na I & 205 & CC\\
Mg I & 232 & CC \\
Al I & 156 & CC \\
Si I & 379 & CC \\
Ca I & 101 & CC and Type Ia\\
Sc II & 136 & CC  \\
Ti I & 69 & CC  \\
V I & 142 & CC and Type Ia \\
Cr I & 170 & CC and Type Ia \\
Mn I & 189 & Type Ia \\
Fe I & 390 & CC and Type Ia  \\
Co I & 173 & HNe \\
Ni I & 303 & Type Ia \\
Cu I & 27 & HNe \\
Zn I & 141 & HNe \\
Y II & 59 & AGB \\
Ba II & 81 & AGB \\
Nd II & 29 & CC \\
Eu II & 20 & CC \\
\hline
\end{tabular}
\end{table}

There are other techniques specifically designed to automatically reduce the dimensionality of a data set. The most popular one certainly is the Principal Component Analysis (PCA; \citealt{Pearson01}). PCA projects the original space of n$_{\rm orig}$ parameters onto a new set of n$_{\rm PCA}\leq$~n$_{\rm orig}$ features which captures the maximum variation in the data set. In other words, while the original data set may contain variables that are not particularly meaningful because they are highly correlated, the PCA finds a new space of parameters where these correlations are minimized. However, capturing the direction of maximum variation in the data set does not necessarily lead to a larger separation of groups in the new space of parameters and, therefore, to better results.  

There are other algorithms for dimensionality reduction that are better suited than PCA for clustering analysis, such as the Linear Discriminant Analysis (LDA; \citealt{Fisher38}). By analysing the distribution of known groups (or classes) of points in the original space of parameters, LDA finds a new set of n$_{\rm LDA}\leq$~n$_{\rm orig}$ features that maximizes the ratio between class variation and within-class variation and that can best characterize all the different classes. As a result, the different classes will be more separated in this new space of parameters. That is exactly what can boost the performances of a clustering algorithm. 

\begin{figure}
\centering
  \includegraphics[width=0.48\textwidth]{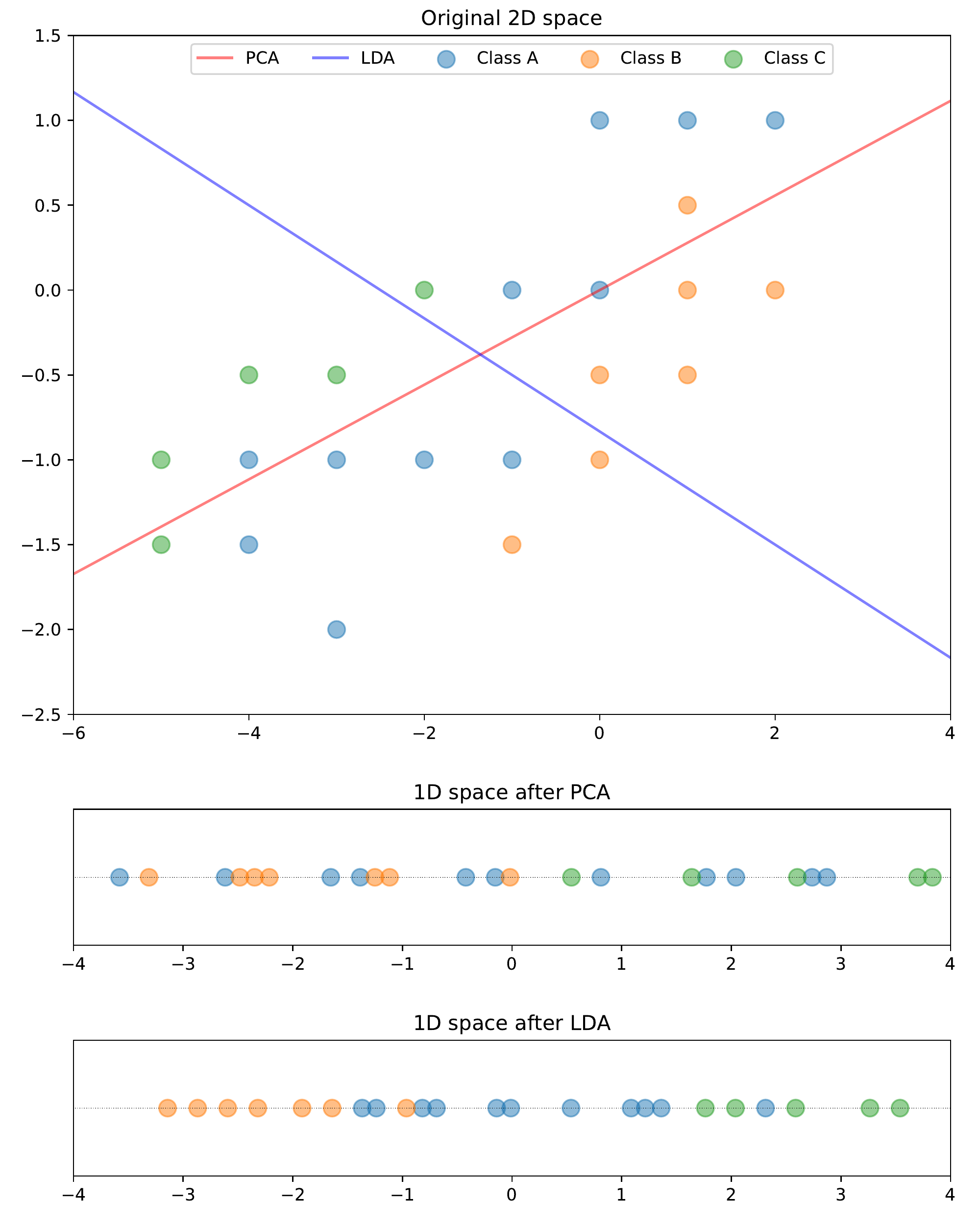}
  \caption{\textbf{Top panel.} Distribution of points belonging to three different classes in a 2D space. The lines represent the projection axis found by PCA (in red) and LDA (in blue). \textbf{Middle panel.} Distribution of points in the 1D space found by the PCA. \textbf{Bottom panel.} Distribution of points in the 1D space found by the LDA.
  }
  \label{Fig:projections}
\end{figure}

The main difference between PCA and LDA is illustrated in the example of Fig.~\ref{Fig:projections}. The top panel shows the distribution of points belonging to three different classes in a 2D space. We also overplot the projection axis found by the PCA (in red) and LDA (in blue). While PCA projects data on the axis with the highest variation, LDA projects on the axis that ensures a better separation of the three classes. The results of the two different approaches for dimensionality reduction are shown in the panels below: while there is a significant overlap between the three classes in the 1D space produced by PCA (mid panel), LDA has found a new space that is much better suited for classification problems and clustering analysis (bottom panel). However, we also stress that there are two possible difficulties in using LDA instead of PCA. First, while PCA is an unsupervised technique, LDA needs to be trained on a set of points whose classes are previously known. In principle this is not always possible. However, in practice, to train LDA here we can use the known members of the open clusters that have been excluded from \textit{optimization} data set. The second difficulty is due to the fact that LDA very easily tends to overfit the data especially when applied on a large number of highly correlated variables. One way to tackle this issue is to use PCA to reduce dimensionality first and then apply LDA only to the principal components. Here we test this latter strategy and we demonstrate that it can further improve the performances of the clustering analysis.

First, we apply the PCA in order to reduce the dimensions of the abundance space from n$_{\rm orig}=19$ to n$_{\rm PCA}=8$. Second, we apply the LDA to further reduce the dimensions to n$_{\rm LDA}=5$. The LDA is trained on the 18 open clusters of the \textit{hidden} data set and that are older than 0.8 Gyr. Note that these open clusters have ages up to 7 Gyr and Galactocentric distances roughly between 6 and 15 kpc. For each of these open clusters we calculate the mean abundances and standard deviations relative to all the 19 elements. Given this information, we randomly drawn 1000 points from the multivariate Gaussian distribution defined for each open cluster in the 19-dimensional abundance space. These 18,000 points are those used to train the LDA. However, before that, we have applied the same standardization and PCA models used on the \textit{optimization} data set. This is necessary in order to ensure that the training set is preprocessed in exactly the same way as the data set what we want to analyse. Once this latter has been transformed by the LDA model, we apply the standard clustering analysis whose results are shown in the bottom panel of Fig.~\ref{Fig:elements}. This technique of reducing the number of dimensions of the abundance space through PCA + LDA outperforms all our previous tests. Now we are able to recover 9 open clusters out of 13. This result is extremely satisfactory, as it also outperforms similar attempts of recovering open clusters in the abundance space \citep[e.g.][]{Casamiquela21}.

\section{Blind chemical tagging in the thin disk}
\label{Sec:chemical_tagging}

The analysis of the \textit{optimization} sample carried out in the previous section has allowed us to determine the best strategies and preprocessing steps that can improve chances of chemical tagging. Instead, in this Section we perform a completely new study using the full data set described in Section~\ref{Sec:dataset}. The overarching aim of this additional analysis is to conduct an experiment that simulates - to the extent possible - a realistic attempt of strong chemical tagging in the thin disk. There is no need to say that all the lessons learnt in the previous Section are applied and exploited here with the objective of achieving higher performances from the clustering analysis. In practice, i) we make use of the OPTICS algorithm, ii) we pre-order and standardize the data set, iii) we use the Manhattan metric, and iv) we reduce the abundance space dimensions with PCA+LDA.

A typical data set that one could use for chemical tagging is composed by a number of known members of open clusters, plus a number of stars with no association to any open cluster. The first serve as a control sample, necessary for tuning the algorithm's hyperparameters  and testing the efficacy of the analysis. Instead, the stars with unknown association include field stars and members of the hidden associations that we wish to recover. Our data set, as it is described in Section~\ref{Sec:dataset}, includes 190 stars from the \textit{optimization} sample which are members of the 13 open clusters. These stars will be our control sample. Furthermore, our data set also includes 112 stars from the \textit{hidden cluster} sample which are members of 23 open clusters. These will be the associations that we will try to identify in the elemental abundance space. These cluster members will be ``diluted'' in a sample of 456 \textit{field stars}, also included in our data set.

Therefore, the goal of this new analysis is to recover as many stellar associations as possible among those of the \textit{hidden cluster} sample. This will be a blind search, as the information on the stellar membership to the open clusters that we wish to recover is never used during the analysis. The only information we are going to use is the one related to the membership of the stars from the \textit{optimization} sample, which will be useful to set the algorithm's hyperparameters.

\subsection{Searching the hyperparameters}
\label{Sec:search_param}

There are three hyperparameters that need to be tuned: \texttt{min\_samples}, n$_{\rm PCA}$, and n$_{\rm LDA}$. They are searched within the following intervals: \texttt{min\_samples}~$\in$~[2..30], n$_{\rm PCA}\in$~[5..15], and n$_{\rm LDA}\in$~[5..n$_{\rm PCA}$]. All the other hyperparameters are set to the values listed in Section~\ref{Sec:hyperparameters}. Before applying the clustering analysis we follow the preprocessing steps described above. Namely, we sort and standardize the data set, we train the LDA on the \textit{optimization} sample and then we apply the dimensionality reduction algorithms PCA + LDA similarly to the procedure described in Section~\ref{Sec:chemical_space}. One significant difference from our previous analysis is that LDA is now trained on the \textit{optimization} sample. The {\sc OPTICS} algorithm is then applied to the whole data set of 767 stars including the \textit{field stars}, the \textit{hidden cluster} sample, and the \textit{optimization} sample. Given the fact that this data set contains four times the number of stars used for the analysis in Section~\ref{Sec:Methods}, here we relax the CE threshold considering all the solutions with CE$\geq$0.05. Therefore, we tune the hyperparameters in order to maximize the number of recovered open clusters among those of the \textit{optimization} sample in a data set which includes also the \textit{field stars} and the \textit{hidden cluster} sample. The hyperparameters satisfying this criterion are the following: \texttt{min\_samples}=3, n$_{\rm PCA}$=11, and n$_{\rm LDA}$=8.

\begin{figure*}
\centering
  \includegraphics[width=0.95\textwidth]{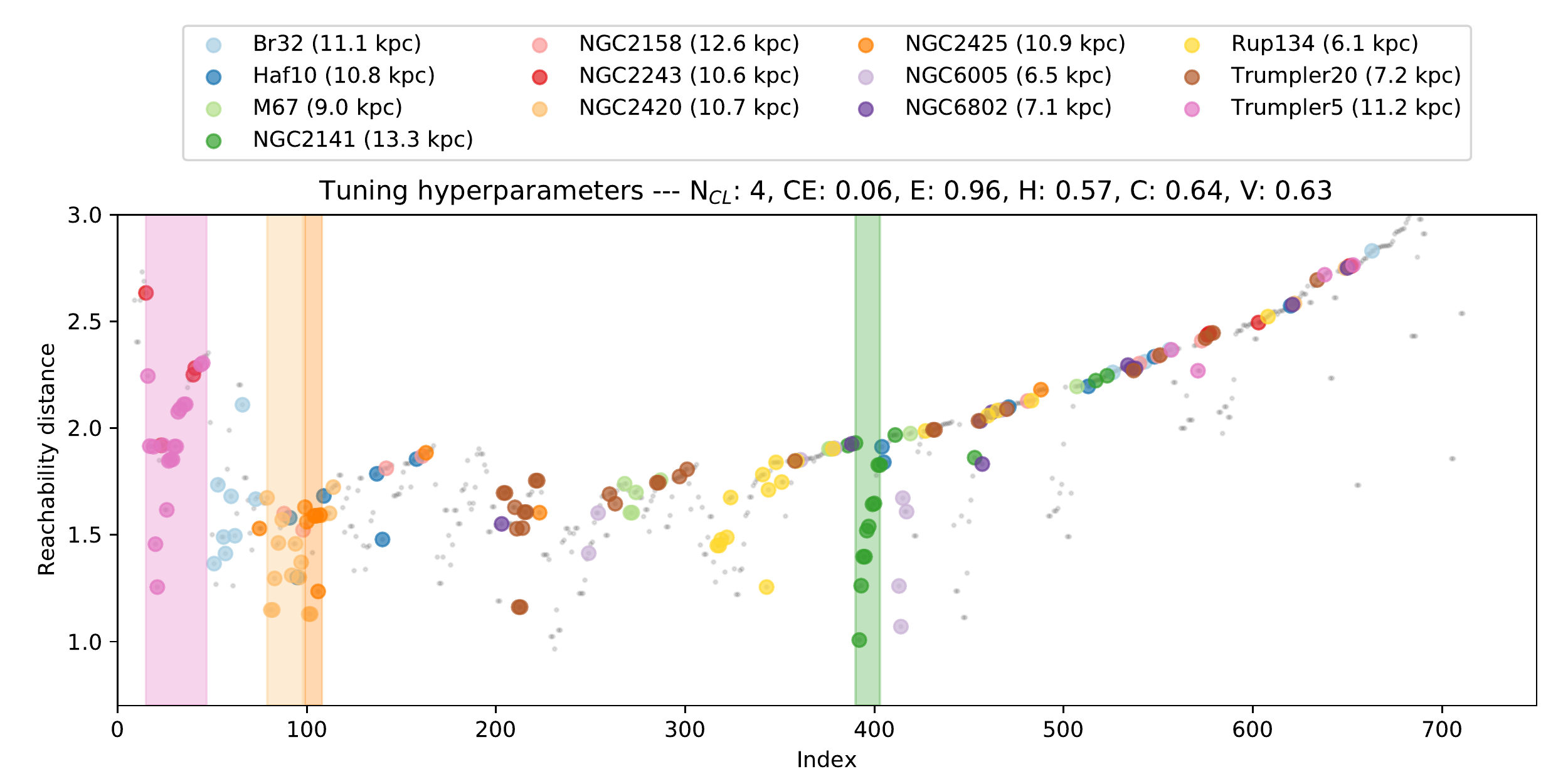}
  \caption{The figure shows the reachability-plot resulting from the clustering analysis of the data set comprising the 190 stars of the \textit{optimizing} sample, 112 stars of the \textit{hidden cluster} sample, and the 465 stars of the \textit{field stars} sample. The open cluster members of the \textit{optimizing} sample are shown with coloured points. All the other stars are represented as gray points. The solution represented here corresponds to the one that maximises the number of recovered clusters from the \textit{optimizing} sample (N$_{\rm cl}$). The overdensities found by {\sc OPTICS} corresponding to the recovered open clusters are highlighted with vertical coloured bands.
  }
  \label{Fig:big_OCs}
\end{figure*}

Specifically, using these hyperparameters we are able to recover four open clusters at H$\geq$0.5 and C$\geq$0.5 from the \textit{optimization} sample. These are NGC~2420, NGC~2141, NGC~2425, and Trumpler~5. The resulting reachability-plot is shown in Fig.~\ref{Fig:big_OCs} from which we can observe that other open clusters from \textit{optimizing} sample  - such as Berkeley~32,  Ruprecht ~134, M~67, and Trumpler~20 - are correctly ordered by {\sc OPTICS} although they are not fully recovered at the H$\geq$0.5 and C$\geq$0.5 threshold. In fact, the inclusion of field stars and members of other open clusters into this data set, which is much larger than that analysed in Section~\ref{Sec:Methods}, is a factor of ''confusion`` for the clustering algorithm, which is thus able to recover a smaller number of open clusters. On the other hand, we also notice that there are dips and valleys filled by stars that are either \textit{field stars} or members of the \textit{hidden cluster} sample. This indicates that the algorithm is able to detect other overdensities among the rest of the data set.

\subsection{Clustering analysis}
\label{Sec:clustering_analysis}

Once we have found the best set of hyperparameters, we proceed with the clustering analysis on the data set exclusively composed by the 456 \text{field stars} plus the 112 stars from the \textit{hidden clusters} sample. In fact, our goal is to recover as many open clusters as possible from those of the \textit{hidden clusters} sample. Therefore, there is no need of further considering the \textit{optimization} sample which is then removed from the analysis.

\begin{figure*}
\centering
  \includegraphics[width=0.9\textwidth]{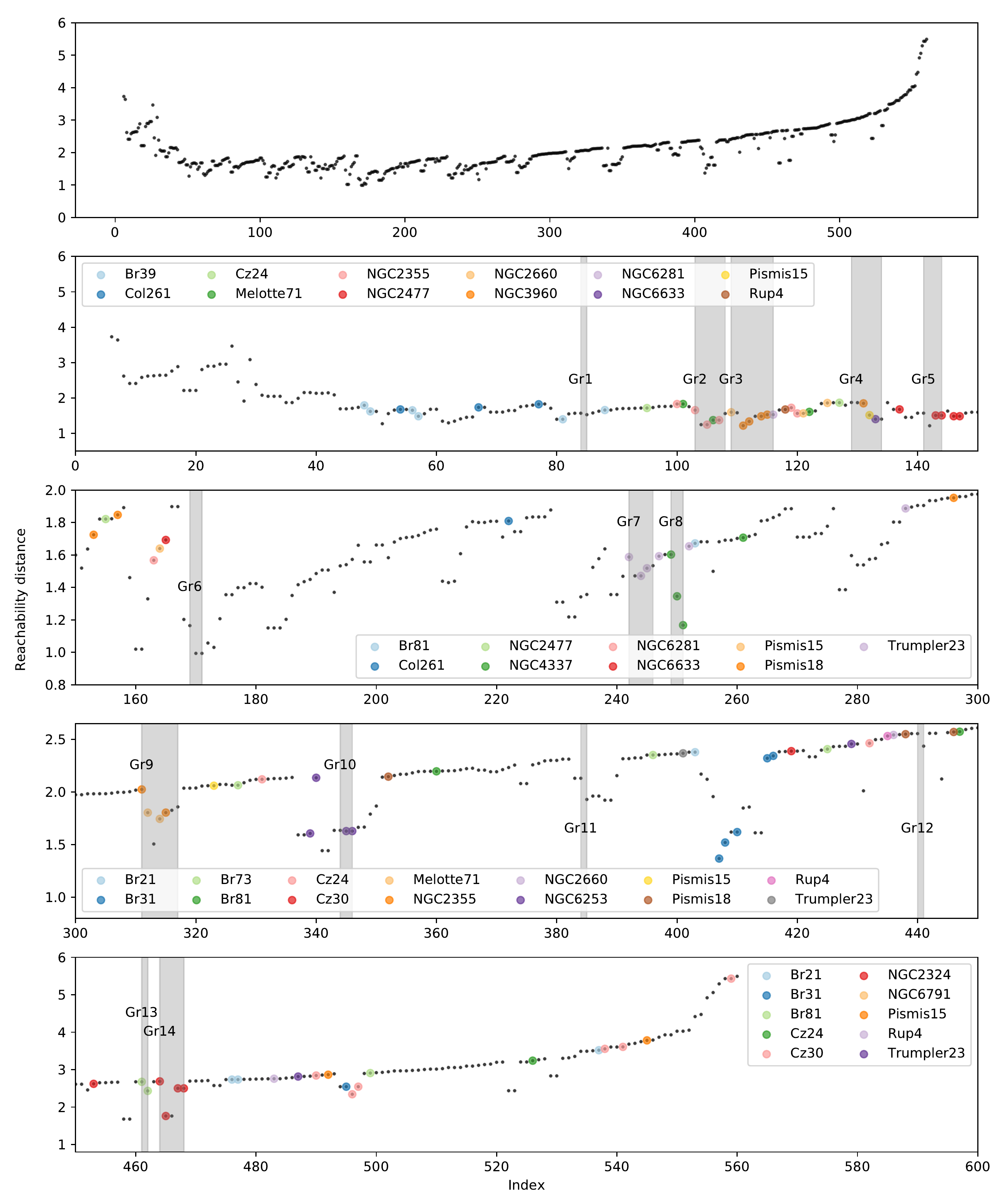}
  \caption{The panels show the reachability plots resulting from the blind clustering analysis of the data set composed by \textit{field stars} and \textit{hidden} open clusters. The top panel shows the entire reachability plot across the full data set. The following panels show smaller portions of the reachability plot of 150 instances each. Stellar members of the 23 open clusters of the \textit{hidden} sample are shown as coloured points. The gray points correspond to the \textit{field stars}. The shaded vertical bands highlight the location of the 14 groups of interest identified as in Section~\ref{Sec:actions}.
  }
  \label{Fig:chem_tagging}
\end{figure*}

The preprocessing steps are the same of Section~\ref{Sec:search_param}, including the LDA which was trained on the 13 open clusters from the \textit{optimizing} sample. Also, the hyperparameres are those found in Section~\ref{Sec:search_param}: \texttt{min\_samples}=3, n$_{\rm PCA}$=11, and n$_{\rm LDA}$=8. The reachability-plot resulting from this analysis is shown as a whole in the top panel of Fig.~\ref{Fig:chem_tagging}, while the four panels below show smaller windows of 150 stars each. In these latter, the actual members of the open clusters we wish to recover are clearly indicated as colored circles. This information is not used in the analysis, but it helps the reader to distinguish the stars we are targeting from the field stars represented as black points. In fact, in Fig.~\ref{Fig:chem_tagging} we notice that stars have been ordered in a way that members of the same open cluster are - in many cases - aligned to adjacent positions. This is mostly evident for open clusters such as Berkeley~31 (index$\sim$45), NGC~3960 (index$\sim$110), Trumpler~23 (index$\sim$200), NGCC~4337 (index$\sim$210), and Berkeley~21 (index$\sim$515). The reachability plot also shows several dips and valleys, indicating the presence of overdensities in the abundance space. As expected, some of these overdensities are mainly formed by members of the same open cluster, such as in the cases of the stellar associations mentioned above. However, we also notice that other dips in the reachability-plot are mostly - or even entirely - composed by field stars. Therefore, our data set may contain other interesting groups of stars characterised by a similar chemistry. On the other hand, it is also possible that these latter groups are just overdensities detected in the noise.

\subsection{Analysis of the overdensity groups found by {\sc OPTICS}}
\label{Sec:actions}

Though our analysis we identify 122 groups of stars clustered in the abundance space. Some of them may be possibly related to real stellar associations. However, there must also be several other groups that are actually spurious clusters due to noise. Furthermore, given that {\sc OPTICS} identifies clusters in a hierarchical structure, some of the groups produced by the algorithm must be too high in this hierarchy and be composed by two or more subclusters. For instance, stellar associations sharing a similar origin within the disk are likely to be very close to each other in the reachability plot and eventually form a unique big valley. Eventually these big valleys are then further divided in smaller dips, which represent the actual open clusters. For instance, although NGC~2420 and NGC~2425 are two individual open clusters, they are enclosed within the same valley in the reachability plots of Fig.~\ref{Fig:projections}. That happens because they share a similar composition, probably because they formed at similar R$_{\rm GC}$ and have similar ages \citep{Cantat-Gaudin20}. However, the aim of our analysis is to identify the smallest groups of stars sharing the same identical origin. We are not aiming at more generalized stellar populations, such as the thin and thick disk, or the inner and outer disk. Therefore, we need some criteria to separate the groups of potential interest from those that are just spurious or not relevant.

In order to do so, we first carry out a selection based on the number of stars included in each groups found by {\sc OPTICS}. Thus, among the initial 122 groups, we consider only those composed by a number of members smaller than 20. This allows us to reject all the groups that are too high in the hierarchy. In fact, given the extent and the nature of our data set, it is extremely unlikely that it contains by chance more than 20 members of a single stellar association. This criterion reduces our sample to 77 groups. A further selection is made by studying how the members within each of the remaining groups are distributed in the space of orbital actions L$_{\rm Z}$-J$_{\rm R}$-J$_{\rm Z}$. Namely, the groups that are denser than a certain threshold will be considered as possible stellar association, while all the others will be discarded.

\begin{figure}
\centering
  \includegraphics[width=0.48\textwidth]{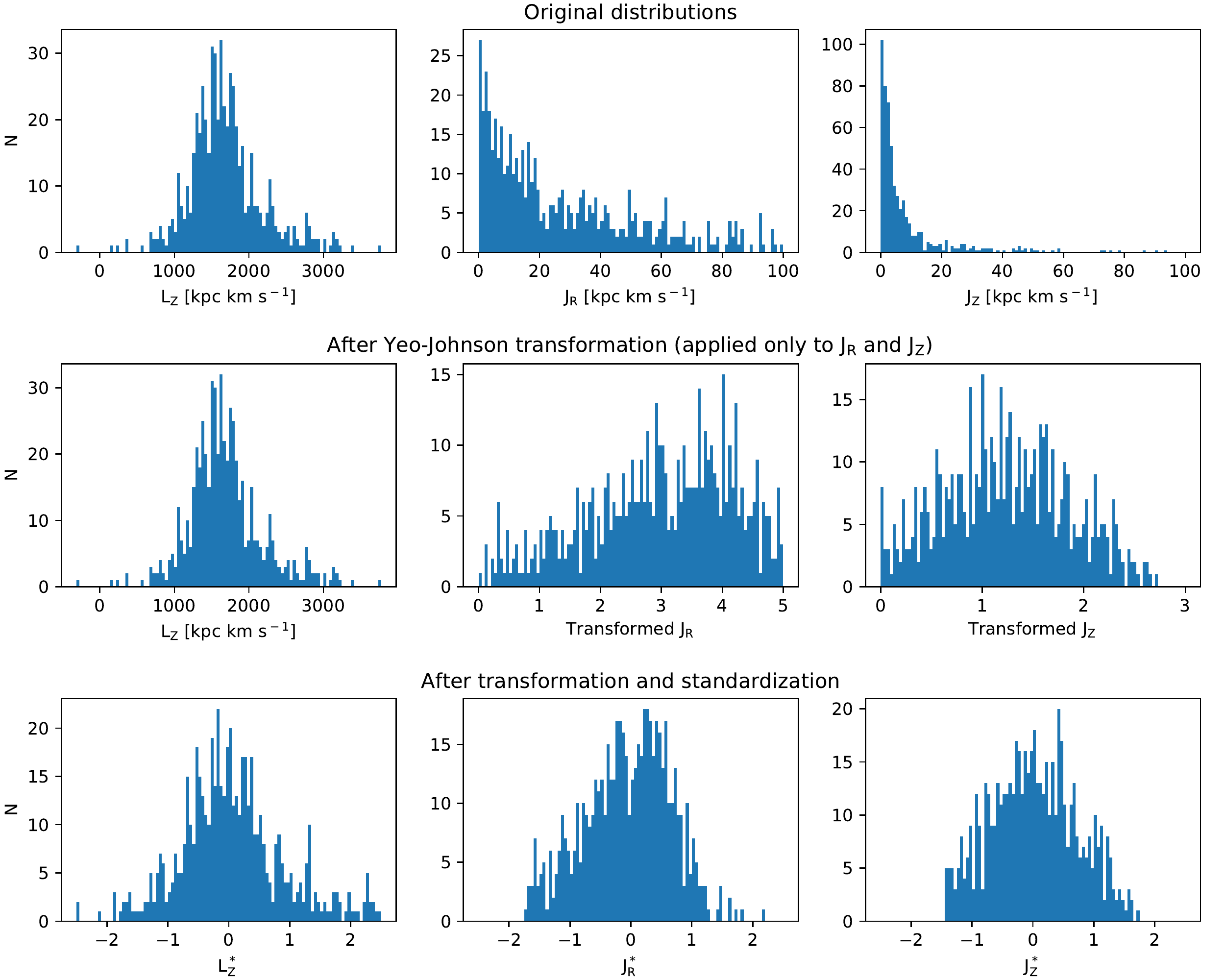}
  \caption{\textbf{Top panels.} Distribution of stars (\textit{field stars} + \textit{optimizing} sample + \textit{hidden cluster} sample) as a function of the orbital actions L$_{\rm Z}$, J$_{\rm R}$, and J$_{\rm Z}$. \textbf{Middle panels.} The orbital action distributions after that the Yeo-Johnson power transformation is applied to the J$_{\rm R}$ and J$_{\rm Z}$ orbital actions. \textbf{Bottom panels.} The orbital actions distributions after the power transformation and standardization.
  }
  \label{Fig:actions}
\end{figure}

A potential issue in studying the dispersion of different groups in the space of actions is due to the fact that the distributions of stars in J$_{\rm R}$ and J$_{\rm Z}$ are strongly peaked at low values (see top panels in Fig.~\ref{Fig:actions}). That poses a problem when using simple measures of spread, such as the standard deviation or the median absolute deviation (MAD), which take a symmetric view of the dispersion giving equal importance to negative and positive deviations from the centre of the distribution. Therefore, in order to alleviate this problem we apply a Yeo-Johnson power transformation \citep{Yeo00} to the J$_{\rm R}$ and J$_{\rm Z}$ variables using the \texttt{PowerTransformer} function from the \texttt{Scikit-learn} {\sc Python} library. As we can observe in Fig.~\ref{Fig:actions}-mid panels, the transformed J$_{\rm R}$ and J$_{\rm Z}$ distributions are now more similar to Normal distributions, although they are not entirely Gaussian. Finally, we standardize the three variables using \texttt{RobustScaler} from \texttt{scikit-learn}. The resulting distributions are shown in Fig.~\ref{Fig:actions}-bottom panels. These new variables define the space of actions L$_{\rm Z}^{*}$-J$_{\rm R}^{*}$-J$_{\rm Z}^{*}$ that we use to discriminate whether a group found by OPTICS contains a potential stellar association or not. 

For each $i-$group we calculate the MADs in the three variables $\{$MAD(L$_{\rm z,i}^*$), MAD(L$_{\rm z,i}^*$), MAD(L$_{\rm z,i}^*$)$\}$. The same quantities are derived for each of the 13 open clusters from the \textit{optimization} sample. Then, we define a metric D$_i$ that we use to discriminate between groups of interest and those that will be rejected, based on the typical internal dispersion that we observe within the open clusters of the \textit{optimization} sample. Namely, D$_i$ is defined as it follows:

\begin{equation}
D_i = \sqrt{\left(\frac{MAD(L_{Z,i}^*)}{T_{L_Z}}\right)^2 + \left(\frac{MAD(J_{R,i}^*)}{T_{J_{R,i}}}\right)^2 + \left(\frac{MAD(J_Z^*)}{T_{J_Z}}\right)^2},
\end{equation}

where T$_{\rm X}$ are threshold values for each action coordinate. They are defined as the average of the MADs calculated for the 13 open clusters in the \textit{optimization} sample plus three times the standard deviation:

\begin{equation}
T_X = \langle\{MAD(X_{OC_j~opt})\}\rangle + 3\times \sigma(\{MAD(X_{OC_j~opt})\}),
\end{equation}

where X$_{OC_j~opt}$ can be one of the coordinates L$_{\rm Z}^{*}$, J$_{\rm R}^{*}$, J$_{\rm Z}^{*}$ observed for the members of the j$^{th}$ open clusters of the \textit{optimization} sample. In other words, we use the $\{$D$_i\}$ metric to rank the dispersion observed in the L$_{\rm Z}^{*}$-J$_{\rm R}^{*}$-J$_{\rm Z}^{*}$ space for each of the 77 groups based on what is observed in the \textit{optimization} sample. All groups with D$_i>$1 are rejected, while the 14 groups with D$_i\leq$1 are considered as groups of interest or potential stellar associations. The location of these 14 groups is indicated in the reachability plots of Fig.~\ref{Fig:chem_tagging} with vertical shaded bands.

The number of \textit{sigmas} used to define T$_{\rm x}$ is arbitrary. A very small number would allow one to recover in the abundance space only the stellar associations that are still physically bound. A very large number would allow one to discover stellar associations that are now fully disrupted, but it would also produce several spurious groups. Here we opt for 3-\textit{sigmas}, which is an intermediate value.

\subsection{The 14 groups of interest}
\label{Sec:groups_of_interest}
Here we provide a detailed analysis of the stellar content of the 14 groups identified through our method. Our main aim is testing whether or not we are able to recover {\em known} clusters hidden among the field stars. We are also interested in identifying eventual escapers or binary members of open clusters. These stars may have been classified as non-members because of kinematics that are discrepant than that of the open cluster. Finally, our data set may also contain disrupted stellar associations that one could potentially recover in the abundance space.

\begin{figure*}
\centering
  \includegraphics[width=0.9\textwidth]{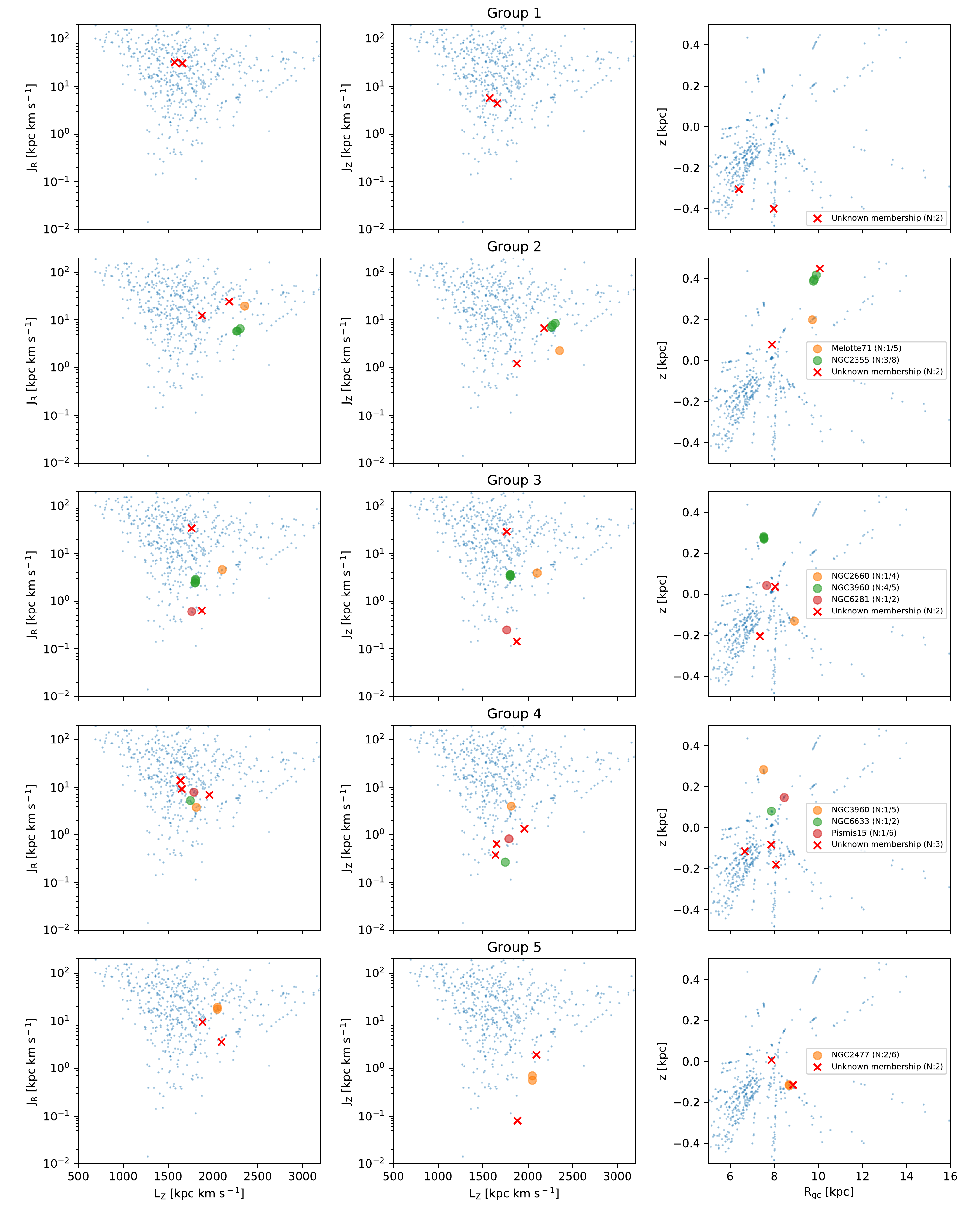}
  \caption{Location of the  groups 1-5 in three planes: J$_R$ versus L$_Z$ (left-side panels), J$_Z$ versus L$_Z$ (central panels), and z versus R$_{\rm GC}$ (right-side panels). 
  }
  \label{Fig:groups_1-5}
\end{figure*}

\begin{figure*}
\centering
  \includegraphics[width=0.9\textwidth]{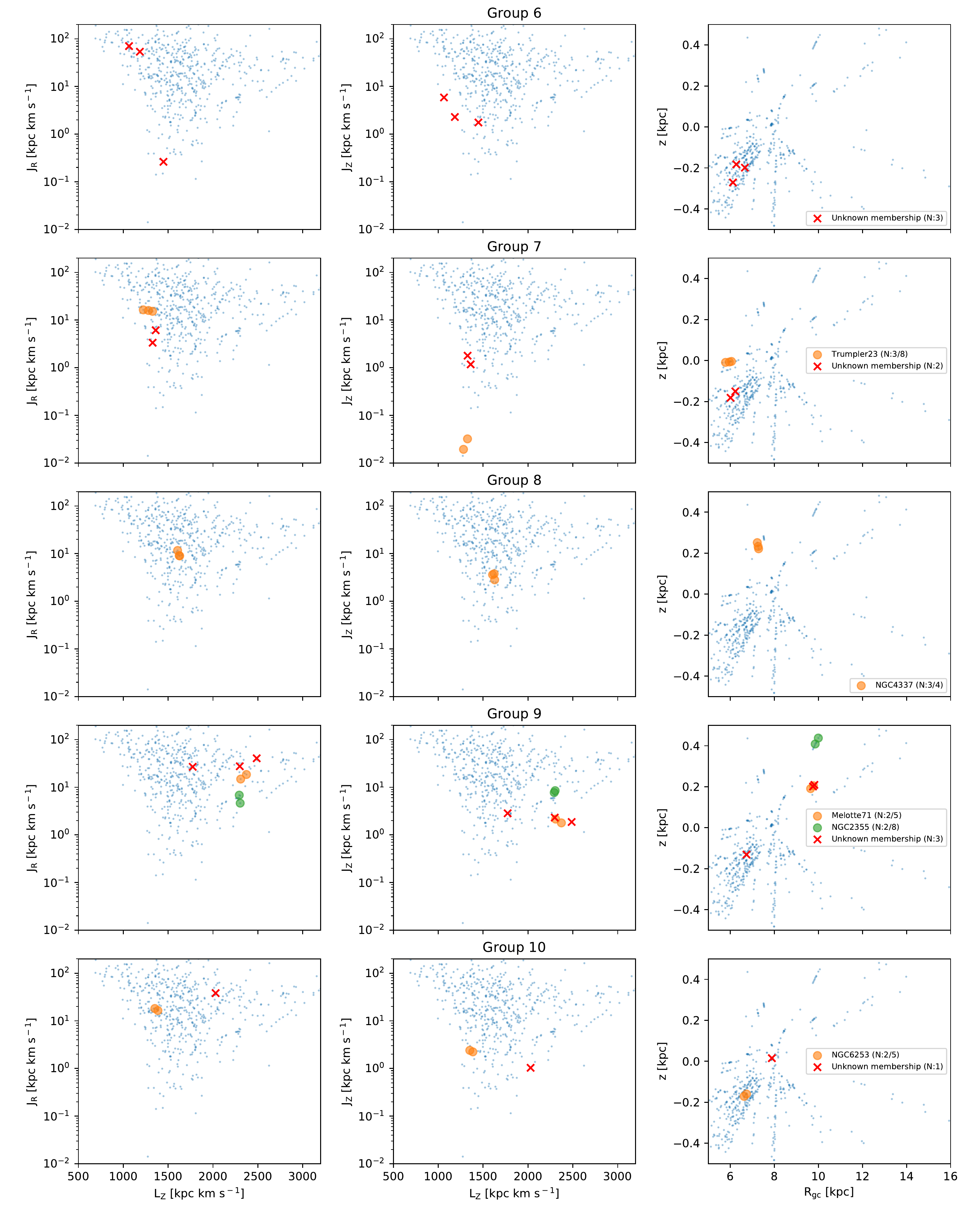}
  \caption{Location of the  groups 6-10 in three planes: J$_R$ versus L$_Z$ (left-side panels), J$_Z$ versus L$_Z$ (central panels), and z versus R$_{\rm GC}$ (right-side panels). 
  }
  \label{Fig:groups_6-10}
\end{figure*}

\begin{figure*}
\centering
  \includegraphics[width=0.9\textwidth]{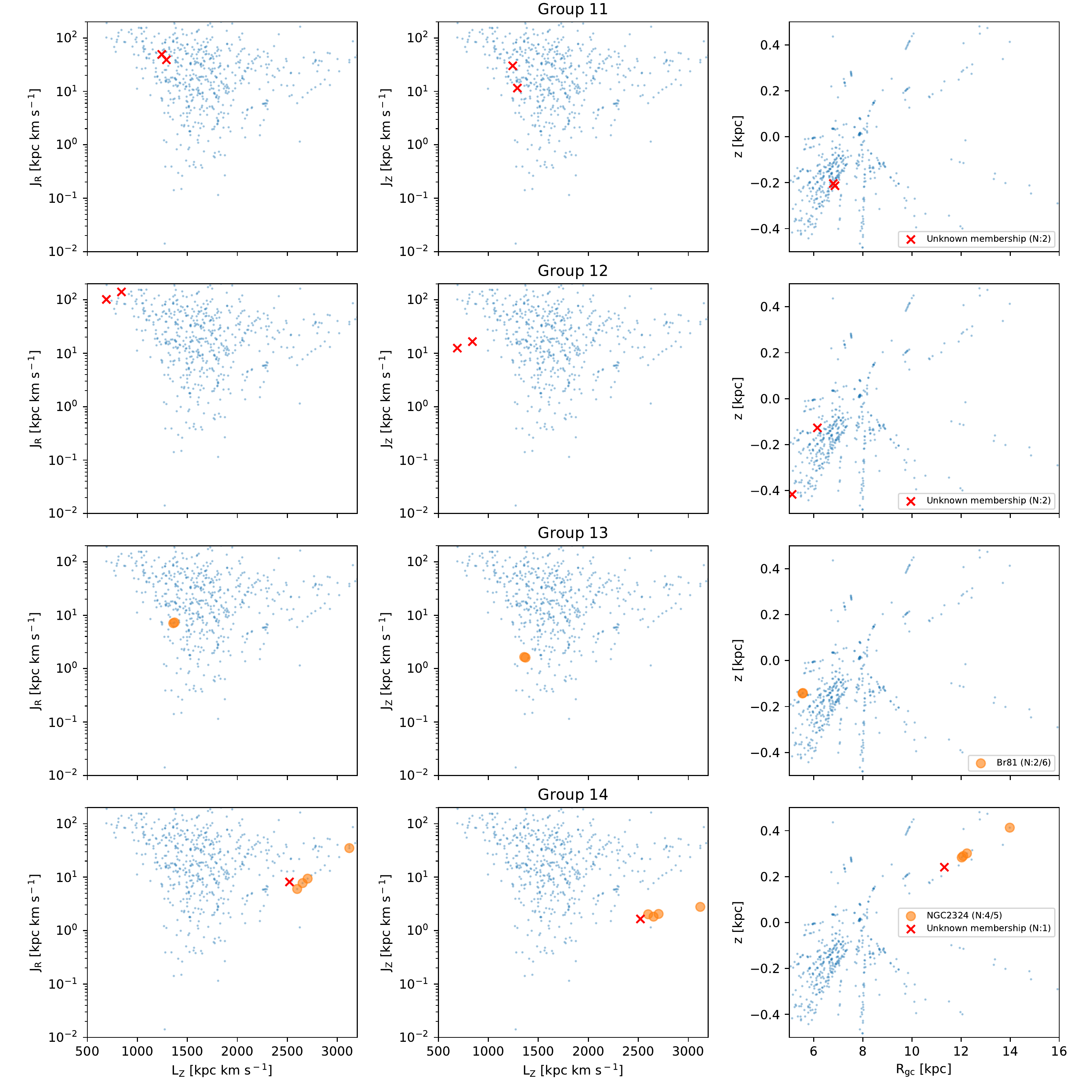}
  \caption{Location of the  groups 11-14 in three planes: J$_R$ versus L$_Z$ (left-side panels), J$_Z$ versus L$_Z$ (central panels), and z versus R$_{\rm GC}$ (right-side panels). 
  }
  \label{Fig:groups_11-14}
\end{figure*}

In Figures~\ref{Fig:groups_1-5}, \ref{Fig:groups_6-10} and \ref{Fig:groups_11-14}, we show the location of the 14 groups in three planes: J$_R$ versus L$_Z$ (left-side panels), J$_Z$ versus L$_Z$ (central panels), and z versus R$_{\rm GC}$ (right-side panels). 
The first two columns of panels tell us the kinematic similarity of the stars, while the last column indicate the spatial relation between them. 

The [C/N] abundance ratio is an excellent proxy of stellar age that can be used for giant stars \citep{Salaris15}. In fact, carbon and nitrogen produced into the stellar interior are brought up to in the stellar surface during the first dredge-up in quantities that depend on the stellar age. In the [C/N] VS age relation calibrated by \citet{Casali19}, the abundance ratios can vary from -1.0 for a 100 Myr old star to -0.2 for a 10 Gyr old star. Therefore, given that carbon and nitrogen have not been used in the clustering analysis, we use this chemical clock when available to provide fundamental and independent information on the real nature of the stellar groups we have identified. The [C/N] ratios are computed from {\em Gaia}-ESO {\sc DR6} abundances as in \citet{Casali19}. We also use their [C/N] VS age relation to estimate the stellar ages. Note that this relation can be used only for stars that have already passed through the first dredge-up.

\paragraph{Group~1} 
It contains two stars, one belonging to one of the CoRoT field and one to the K2.  Although they share the same chemical composition, the two stars are not spatially close, being about 2 kpc apart. No information about their ages, nor [C/N] is available. 


\paragraph{Group~2} 
It contains four members, one in Melotte~71 and three in NGC~2355. The two clusters are similar for R$_{\rm GC}$, close to 10~kpc, and  age. The group contains also two field stars. 
Interestingly, the star with CNAME=07162986+1344207 is in the field of NGC~2355. Although it has radial velocity (70.9$\pm$0.4 km s$ ^{-1}$) and proper motions ($\mu_\alpha$=-0.665$\pm$0.014 mas/yr, $\mu_\delta$=-0.324$\pm$0.011 mas/yr) that are different from those of NGC~2355 (36$\pm$1 km s$ ^{-1}$, -3.80$\pm$0.0.14  mas/yr, and -1.09$\pm$0.0.13  mas/yr; \citealt{Cantat-Gaudin20}), the parallax is consistent within the uncertainties (0.486$\pm$0.012 mas). In addition, the star has a [C/N]=-0.54$\pm$0.10 dex that is consistent with that of the open cluster (i.e., -0.51$\pm$0.04 dex). All these features combined with their chemical similarity suggest that 07162986+1344207 is a true member of NGC~2355 which is escaping from the open clusters. The star was disregarded as part of the open cluster due to the discrepant kinematics, however we assign it to NGC~2355 through {\em strong} chemical tagging.

\paragraph{Group~3} 
It is composed by a mixture of members of NGC~2660, NGC~3960, NGC~6281 and two field stars. 
The NGC~3960 cluster is recovered within this group with completeness C=0.80, homogeneity H=0.50 and V-value V=0.62.
The three clusters whose members have been identified are located at R$_{\rm GC}$ from 7.7 to 8.9~kpc and with log(age) from 8.7 to 9 \citep{Cantat-Gaudin20}.


\paragraph{Group~4} 
Members of three clusters with similar ages (NGC~3960, NGC~6633, Pismis15), but a slightly different location in the disc (R$_{\rm GC}$=7.6, 8.6, 8.0~kpc). In addition, the group contains three field stars. There is not an evident kinematic connection between all these objects.


\paragraph{Group~5} 
We have correctly re-identified two members of NGC~2477, together with other two field stars. These latter do not show any evident connection to NGC~2477.


\paragraph{Group~6} 
It contains three stars located in the inner disc, towards the bulge. They have similar R$_{\rm GC}$ and z. Two of them (CNAMEs 18101550-3147076 and 18161182-3339161) are also very similar in their actions and they have similar abundance ratios between carbon and nitrogen, [C/N]=-0.50$\pm$0.04 and [C/N]=-0.61$\pm$0.03. They can be candidate member of a possible common site of star formation. The third star has a very different [C/N]=-0.27$\pm$0.06.

\paragraph{Group~7} 
We have re-identified three members of Trumpler~23. The group also contains two field stars located in the inner disc, towards the bulge. These latter share the same R$_{\rm GC}$ of Trumpler~23, but different z. 
Between them the two field stars have similar height in the plane and kinematic properties (possible common site of formation). 

\paragraph{Group~8} 
We have re-identified three members (out of 4) of NGC~4337. The cluster is recovered with C=0.75, H=1.0, and V=0.86.

\paragraph{Group~9} 
The group contains members of Melotte~71 (2/5) and of NGC~2353 (2/8). It also contains two stars in the field of Melotte 71 (possible escapers for evaporation or binary stars, with difference in radial velocity within 20~km s$^{-1}$ from the mean cluster velocity) and  an interesting star in the inner disc, with the same composition, but without any kinematic connection with the clusters' members. 
The radial velocities of the two stars in the field of Melotte 71 measured in {\em Gaia}-ESO {\sc idr6} and in {\em Gaia} {\sc edr3}, thus at two different epochs, indicate variations, favouring the hypothesis of binarity. Among the two candidate members of Melotte 71 - CNAMES 07372506-1202241 and 07373561-1202542 - only the first has a determination of [C/N]=-0.73$\pm$0.09, which is consistent with that of the open cluster, [C/N]=-0.63$\pm$0.07.

\paragraph{Group~10} 
The group contains two members (2/5) of NGC~6253. The cluster is recovered within this group at C=0.40, H=0.67, and V=0.50. The group also includes one field star that is spatially and kinematically unrelated to NGC~6253.

\paragraph{Group~11} 
It is composed by two stars in one of the CoRoT fields observed by {\em Gaia}-ESO, which share similar  kinematic properties, location and chemistry. They may have originated from the same star-formation site. 

\paragraph{Group~12} 
The group contains two field stars with similar R$_{\rm GC}$, but different z. 

\paragraph{Group~13} 
The group contains two members (2/8) of Berkeley~81. The cluster is recovered within this group at C=0.33, H=1.0, and V=0.5.

\paragraph{Group~14} 
We have re-identified four members of NGC~2324. The cluster is recovered within this group at C=0.80, H=0.80, and V=0.80. The group also includes one star (CNAME=07040822+0105185) located in the field of NGC~2324. It has a radial velocity 34.9$\pm$0.4 km s$^{-1}$, and proper motion ($\mu_\alpha$=-0.27$\pm$0.02 mas/yr, $\mu_\delta$=-0.06$\pm$0.02 mas/yr) that are consistent with those of the open cluster (-0.31$\pm$0.11 mas/yr, and -0.09$\pm$0.09 mas/yr). Only parallax is marginally inconsistent: 0.0284$\pm$0.018 mas for the star and 0.19$\pm$0.06 mas for the cluster. Unfortunately the [C/N] ratio is unavailable. However, the high similarity in kinematics strongly suggests that 07040822+0105185 is a member of NGC~2324 that we recovered through {\em strong} chemical tagging.

\hfill

Finally, some general remarks on the nature of the groups identified:

{\em i)} the chemical homogeneity of each group is in the range 0.01-0.1 dex. If there are members of known star clusters, they are usually more homogeneous than the other stars; {\em ii)} in some cases, groups contain members of stars in known clusters and field stars, which are chemically similar, but there are no evident connections in terms of their location in the disc and kinematic properties;  {\em iii)} there are some groups (or sub-groups within groups) composed by {\sc GE\_BL} stars (meaning stars located towards the bulge, in the inner disc). They share chemistry and often also R$_{\rm GC}$, being possible candidate members of disrupted clusters. Although this result is important, we must also remember that we have a bias towards the inner disc when selecting field giants (see group 6, and sub-group 7) because of the {\em Gaia}-ESO selection function\citep{Stonkute16}; {\em iv)} groups with members of more than one cluster usually contain members of clusters located at the same R$_{\rm GC}$. This may be due to a bias introduced the metric D used to select the groups of interest. However, it may also indicate the major role played by the position in the disc for chemical evolution; {\em v)} we find stars in the field of clusters which are not classified as members, but whose chemical composition is similar to the other members (see groups 2, 9, and 14). They can be binary stars (SB1) or cluster {\em escapers} \citep{moyano13}, {\em vi)} we find a possible signature of a disrupted cluster in one of the CoRoT fields (group 11).

\section{Summary and conclusions}
\label{Sec:conclusions}

The goal of this study is performing an experiment on strong chemical tagging, testing whether or not we are capable of blindly recovering members of open clusters through clustering analysis in the abundance space. For this experiment we use a controlled sample of giant stars observed with UVES by the {\em Gaia}-ESO survey. The sample contains both members of open clusters and field stars. 

\subsection{Practical tips for clustering in the elemental abundance space}

The success of clustering analysis relies on some particular decisions that the scientist has to take about the  pre-processing techniques and the hyper-parameters' values. Unfortunately, there is no rule of thumb, nor a general strategy that works well for all types of problems. Instead, the correct strategy strongly depends on the type of problem one has to tackle and the type of data set one has at hand. Therefore, as a first step of our analysis, we carry out a preliminary study specifically designed to explore different strategies of analysis. The aim is to identify the best procedure for recovering members of open clusters within our data set. For this study we use a sample of 13 different open clusters, containing at least 10 members each. Given its function in our work, the sample is called \textit{optimization} sample and it comprises 190 stars in total. From the preliminary study we learn a few lessons that are valid for any chemical tagging problem:
\begin{itemize}
    \item Open clusters have different densities in abundance space. The algorithms already implemented in \texttt{Python} that can process a data set containing both clusters and noisy data and that can capture groups of different densities are {\sc OPTICS} and {\sc HDBSCAN}. Specifically, we decide to use {\sc OPTICS} because of its simplicity and versatility (it has a smaller number of hyper-parameters to be tuned than {\sc HDBSCAN}) and also because there are no chemical tagging experiments with {\sc OPTICS} in the literature yet.
    \item The distance- and density-based algorithms - such as k-means, {\sc DBSCAN, HDBSCAN} and {\sc OPTICS} - are not entirely deterministic. Their outcomes slightly depend on how the data set is initially ordered. Therefore, with the aim of boosting the performances of the clustering algorithms and of ensuring the reproducibility of our results, we pre-order our dataset sorting as a function of chemical abundances, stating with [Fe/H].
    \item Standardization is always recommended as it gives all dimensions an equal weight.
    \item The Manhattan metric should be preferred over the Euclidean metric. This is especially important if we are working with a noisy data set and on a high-dimensional space. 
    \item Using all the chemical abundances that one has at hand is a strategy that does not pay. Instead, the clustering analysis should be carried out using only the most relevant features. The choice of the best elements for chemical tagging depends on the chemical information they carry, but also on the typical precision of their abundance measurements. Therefore, there we expect that the best set of elements could vary from survey to survey. The most relevant features can be chosen by the scientist using criteria such as i) the nucleosynthetic history of each specific element, and ii) the Calinski-Harabasz score. There are other methods for an automatic dimensionality reduction, such as the PCA, the LDA, and many others. 
    \item PCA alone is not the most convenient strategy to reduce the dimensionality for chemical tagging problems. This is demonstrated in Fig.~\ref{Fig:projections}. Instead, better results are expected by applying in sequence PCA and LDA, where LDA can be trained on a set of known cluster members.
\end{itemize}

By putting these lessons into practice we are able to recover 9/13 open clusters into groups at the 50$\%$ of completeness and homogeneity. This result outclass previous studies attempting non-blind chemical tagging on samples solely composed by open clusters' members \citep[e.g.][]{Blanco-Cuaresma18,GarciaDias19,Casamiquela21}.

\subsection{Blind chemical tagging}

One of the main goals of this study is testing whether clustering analysis in the abundance space can recover the building blocks of the thin disk. To do this, we carry out an experiment of blind chemical tagging aiming at the identification of open clusters hidden in our data set. 

Thus, following the strategy outlined above, we analyse the data set composed by 456 \textit{field stars} and 112 members of 23 open clusters. These latter are the \textit{hidden clusters} that we aim to recover in the abundance space. The analysis is completely ``blind'', meaning that we never use information of the stellar membership to reconstruct the \textit{hidden clusters}. In fact, the algorithm's hyperparameters are optimised exclusively on the \textit{optimisation} sample. Nevertheless, the actual membership of each single star from the \textit{hidden clusters} sample is known and is used to evaluate our results. The clustering analysis identifies 122 overdensities in the abundance space. We consider only the groups with a number of stars ranging between 2 and 20 and with a dispersion in the space of orbital actions below a certain threshold. We think that these 14 groups are those with the highest chances of coinciding with some of the \textit{hidden clusters}.

Interestingly three of the 14 groups coincide with open clusters recovered at the C$\geq$0.5 and H$\geq$0.5 threshold. These are the groups $\#$3, $\#$8, and $\#$14. The open clusters are NGC~3960, NGC~4337, and NGC~2324, respectively. There are other two groups coinciding with open clusters recovered at the threshold of V$\geq$0.5. These are the group $\#$10 for NGC~6253 and the group $\#$10 for Berkeley~81. In addition, we identify four stars that are possibly escapers or binary members of NGC~2355 (group $\#$2), Melotte~71 (group $\#$9), and NGC~2324 (group $\#$14). These stars were initially classified as field stars, due to the marginal discrepancy between their kinematics and that of their open cluster. 

In conclusion, based on the knowledge that is currently available to us, 7/14 groups carry real information around stars formed within the same stellar association. The other seven groups candidate stellar associations which should be further investigated with follow-up observations.

\subsection{The {\sc OPTICS} execution time}

The algorithm execution time is a very important aspect of data clustering. This is especially true when the data set is composed by a large number of instances and features. Our data set is extremely small and the clustering analysis described in Section~\ref{Sec:clustering_analysis} (i.e., 568 instances and 8 features) has taken 0.4 seconds with a 3.1 GHz dual-core processor. 

Given that the time complexity\footnote{Time complexity is the amount of time taken by an algorithm to run, as a function of the length of the input.} of the {\sc OPTICS} algorithms scales with the square of the number of instances, a data set of 10,000 instances would take around 20 seconds. Such a data set is more similar to those that could be employed by future spectroscopic surveys. Tuning the hyper-parameters under these conditions - e.g., running the algorithms $\sim$100 times - would be still feasible with our computing resources. Instead, a data set composed by 100,000 instances would easily take around 30 minutes to run and would require a much faster machine.

However, the {\sc OPTICS} algorithm includes the \texttt{max$\_$eps} hyper-parameter which describes the maximum distance between two points for one to be considered as in the neighborhood of the other and that can be tuned in order to reduce the execution time. The \texttt{max$\_$eps} hyper-parameter is set to infinity by default. However, reducing its value reduces the volume considered around each point to find its neighbours, thus it can also speeds up the execution time. 

\subsection{Concluding remarks}
{\em Strong} chemical tagging - if fully feasible - would allow us to disentangle the spatial distribution of stars from their radial migration within the disk and would allow us to reconstruct the building blocks of the thin disk. The results outlined above are very promising for the prospect of using chemical abundances to tag stars to their birth clusters. On the other hand, we are still far from being fully satisfied, as the recovery fraction of open clusters in the blind analysis tends not to be particularly high. It is still unclear whether this recovery fraction is intrinsically low due to the small chemical diversity of stellar associations within the thin disk, or instead it is due to other limiting factors that we can overcome with a more performing algorithm and more precise data.


One should also notice that the data set used in our experiment is strongly unbalanced compared to what one would expect under real circumstances. In fact, about 1/4 of the stars in our data set are members of the \textit{hidden clusters}, which has certainly made easier the recovery of open clusters. Nevertheless, open clusters have been recovered, as well as new members that were lost from the membership analysis. This certainly leaves a door open to feasibility of strong chemical within the thin disk.

In this regard, one should also consider which is the prospect of stellar astronomy and data science for the next decade. The forthcoming large spectroscopic surveys and facilities (e.g., 4MOST, WEAVE, MOONS; \citealt{Cirasuolo14,Dalton16,deJong19}) are expected to provide huge data sets, but - given their limited spectral resolution - they will probably provide abundance determinations that are similar or even more noisy than those used here. Likely, this is a big limitation for future attempts of chemical tagging. However, it is conceivable that after 2030 we will see the first large high-resolution spectroscopic survey of our Galaxy, which will increase our chances of finally using this technique. 

Finally, one should not forget that data science is a discipline that is currently progressing at high speed. This momentum is due to the constant acquisition of data from states, multinationals, industries, etc... It is a practice that will further increase during the next years. Thus, the current techniques of clustering analysis will be certainly improved and other algorithms will be developed. For instance, new techniques of clustering analysis are being developed to deal with noise, outliers, and data sets with missing values \citep{Song21,Yan21}. These are the same limitations that cripple our efforts of chemical tagging. Therefore, we must not lose hope of effective and complete chemical tagging in our Galaxy: the refinement of techniques combined with the big databases that will be available in the next future, will allow us to study the star formation history of the Galactic disk in great detail.


\begin{acknowledgements}
Based on data products from observations made with ESO Telescopes at the La Silla Paranal Observatory under programme ID 188.B-3002. These data products have been processed by the Cambridge Astronomy Survey Unit (CASU) at the Institute of Astronomy, University of Cambridge, and by the FLAMES/UVES reduction team at INAF/Osservatorio Astrofisico di Arcetri. These data have been obtained from the Gaia-ESO Survey Data Archive, prepared and hosted by the Wide Field Astronomy Unit, Institute for Astronomy, University of Edinburgh, which is funded by the UK Science and Technology Facilities Council.
This work was partly supported by the European Union FP7 programme through ERC grant number 320360 and by the Leverhulme Trust through grant RPG-2012-541. We acknowledge the support from INAF and Ministero dell' Istruzione, dell' Universit\`a' e della Ricerca (MIUR) in the form of the grant "Premiale VLT 2012" and "Premiale 2016 MITiC". The results presented here benefit from discussions held during the Gaia-ESO workshops and conferences supported by the ESF (European Science Foundation) through the GREAT Research Network Programme.
This work has made use of data from the European Space Agency (ESA) mission {\it Gaia} (\url{https://www.cosmos.esa.int/gaia}), processed by the {\it Gaia} Data Processing and Analysis Consortium (DPAC, \url{https://www.cosmos.esa.int/web/gaia/dpac/consortium}). Funding for the DPAC has been provided by national institutions, in particular the institutions participating in the {\it Gaia} Multilateral Agreement. 
LS is supported by the Italian Space Agency (ASI) through contract 2018-24-HH.0 to the National Institute for Astrophysics (INAF).  LM thank the COST Action CA18104: MW-Gaia. GC acknowledges support from the European Research Council Consolidator Grant funding scheme (project ASTEROCHRONOMETRY, G.A. n. 772293, \url{http://www.asterochronometry.eu}). TB was funded by the project grant No. 2018-04857 from the Swedish Research Council.
\end{acknowledgements}

%
%
\bibliographystyle{aa}
\bibliography{Bibliography} 

\end{document}